\documentclass[preprint,amsmath,amssymb,aps,pra,nofootinbib]{revtex4-1}

\usepackage{graphicx,mathptmx}
\usepackage{hyperref}
\usepackage{cleveref}

\hypersetup{colorlinks=true,citecolor=blue,urlcolor=blue}

\newcommand{\Wcmsqd}{\mathrm{W}\text{cm}^{-2}}
\newcommand{\rmd}{\mathrm{d}}
\newcommand{\micron}{{\mu\mathrm{m}}}
\newcommand{\Power}{\mathcal{P}}
\newcommand{\Energy}{\mathcal{E}}

\newcommand{\fs}{\mathrm{fs}}

\newcommand{\abs}[1]{\left| #1 \right|}
\newcommand{\avg}[1]{\langle #1 \rangle}
\newcommand{\Ecrit}{E_\text{cr}}

\newcommand{\um}{u^-}
\newcommand{\up}{u^+}
\renewcommand{\vec}{\mathbf}

\begin{document}

\title{Radiation reaction in electron-beam interactions with high-intensity lasers}

\author{T. G. Blackburn}
\email{tom.blackburn@physics.gu.se}
\affiliation{Department of Physics, University of Gothenburg, Gothenburg SE-41296, Sweden}

\date{\today}

\begin{abstract}
Charged particles accelerated by electromagnetic fields emit radiation, which must, by the
conservation of momentum, exert a recoil on the emitting particle. The force of this recoil,
known as radiation reaction, strongly affects the dynamics of ultrarelativistic electrons
in intense electromagnetic fields. Such environments are found astrophysically, e.g. in
neutron star magnetospheres, and will be created in laser-matter
experiments in the next generation of high-intensity laser facilities.
In many of these scenarios, the energy of an individual photon of the radiation can
be comparable to the energy of the emitting particle, which necessitates modelling
not only of radiation reaction, but \emph{quantum} radiation reaction.
The worldwide development of multi-petawatt laser systems in large-scale facilities,
and the expectation that they will create focussed electromagnetic fields with
unprecedented intensities $> 10^{23}~\Wcmsqd$, has motivated renewed interest
in these effects.
In this paper I review theoretical and experimental progress towards understanding
radiation reaction, and quantum effects on the same, in high-intensity laser fields
that are probed with ultrarelativistic electron beams.
In particular, we will discuss how analytical and numerical
methods give insight
into new kinds of radiation-reaction-induced dynamics,
as well as how the same physics can be explored in experiments at currently existing laser
facilities.
\end{abstract}

\maketitle

\tableofcontents

\section{Introduction}
\label{sec:Parameters}

It is a well-established experimental fact that charged particles,
accelerating under the action of externally imposed electromagnetic
fields, emit radiation~\citep{lienard.ee.1898,wiechert.an.1900}.
The characteristics of this
radiation depend strongly upon the magnitude of the acceleration
as well as the shape of the particle trajectory. For example, if
relativistic electrons are made to oscillate transversely by a field
configuration that has some characteristic frequency $\omega_0$,
they will emit radiation that has characteristic frequency
$2 \gamma^2 \omega_0$, where $\gamma$ is their Lorentz factor.
Given $\omega_0$ corresponding to a wavelength of one micron and an
electron energy of order 100~MeV, this easily approaches the 100s
of keV or multi-MeV range~\citep{corde.rmp.2013}.

The total power radiated, as we shall see, increases strongly with
$\gamma$ and the magnitude of the acceleration.
We can then ask: as radiation carries energy and momentum, how do we
account for the recoil it must exert on the particle?
Equivalently, how do we determine the trajectory when one electromagnetic
force acting on the particle is imposed externally and the other arises
from the particle itself?
That this remains an active and interesting area of research is a
testament not only to the challenges in measuring radiation reaction
effects experimentally~\citep{samarin.jmo.2017}, but also to the
difficulties of the theory itself~\citep{dipiazza.rmp.2012,burton.cp.2014}.
The `correct' formulation of radiation reaction within classical
electrodynamics has not yet been absolutely established, nor has the complete
corresponding theory in quantum electrodynamics. While these points
are undoubtedly of fundamental interest, it is important to note that
radiation reaction and quantum effects will be unavoidable
in experiments with high-intensity lasers and therefore these questions
are of immense practical interest as well.

This is motivated by the fast-paced development of large-scale,
multipetawatt laser facilities~\citep{danson.hplse.2019}:
today's facilities reach focussed intensities of order
$10^{22}~\Wcmsqd$~\citep{bahk.ol.2004,sung.ol.2017,kiriyama.ol.2018},
and those upcoming, such as Apollon~\citep{papadopoulos.hpl.2016},
ELI-Beamlines~\citep{weber.mre.2017} and Nuclear Physics~\citep{gales.rpp.2018},
aim to reach more than $10^{23}~\Wcmsqd$, with the added
capability of providing multiple laser pulses to the same target
chamber.
At these intensities, radiation reaction will be comparable in
magnitude to the Lorentz force, rather than being a small correction,
as is familiar from storage rings or synchrotrons.
Furthermore, significant quantum corrections to radiation reaction
are expected~\citep{dipiazza.rmp.2012}, which profoundly alters
the nature of particle dynamics in strong fields.

The purpose of this review is to introduce the means by
which radiation reaction, and quantum effects on the same,
are understood, how they are incorporated into numerical
simulations, and how they can be measured in experiments.
While there is now an extensive body of literature considering
experimental prospects with future laser systems,
our particular focus will be the relevance to \emph{today's}
high-intensity lasers.
It is important to note that much of the same physics
can be explored by probing such a laser with an ultrarelativistic
electron beam. Previously such experiments demanded a
large conventional accelerator~\citep{bula.prl.1996,burke.prl.1997},
but now `all-optical' realization of the colliding beams
geometry is possible thanks to ongoing advances in
laser-wakefield acceleration~\citep{esarey.rmp.2009,bulanov.nima.2011}.
Indeed, the first experiments to measure radiation-reaction
effects in this configuration have recently been
reported by \citet{cole.prx.2018,poder.prx.2018}.
This review attempts to provide the theory context for the
interest in their results.

Let us begin by introducing
the various parameters that determine the importance of
radiation emission, radiation reaction, and quantum effects.
We work throughout in natural units such that
the reduced Planck's constant $\hbar$, the speed of light $c$
and the vacuum permittivity $\varepsilon_0$ are all equal to
unity: $\hbar = c = \varepsilon_0 = 1$. In these units the
fine-structure constant $\alpha = e^2/(4\pi)$, where $e$ is
the elementary charge.

\begin{figure}
	\centering
	\includegraphics[width=0.6\linewidth]{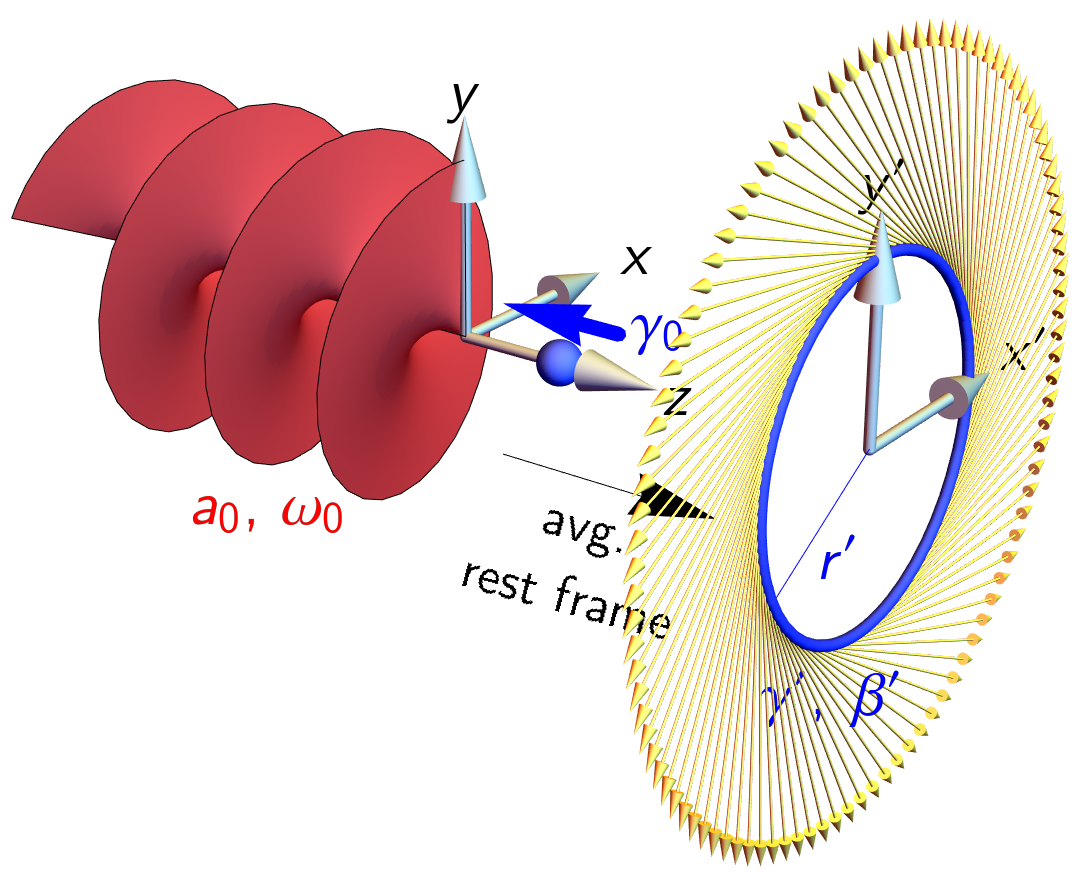}
	\caption{
		Interaction of an electron (initial Lorentz factor
		$\gamma_0$) and a circularly polarized electromagnetic
		wave (frequency $\omega_0$ and normalized
		amplitude $a_0$).
		In its average rest frame the electron
		is accelerated on a circular trajectory, with Lorentz factor
		$\gamma' = (1 + a_0^2)^{1/2}$, velocity $\beta'$ and
		radius $r'$.
		The acceleration leads to the emission of synchrotron
		radiation, which has	characteristic frequency
		$\omega' \simeq \gamma'^3/r'$.}
	\label{fig:LaserSynch}
	\end{figure}

It will be helpful to consider the concrete example shown in
\cref{fig:LaserSynch}. Here an electron is accelerated by a circularly
polarized, monochromatic plane electromagnetic wave. The wave has
angular frequency $\omega_0$ and dimensionless amplitude
$a_0 = e E_0 / (m \omega_0)$, where $E_0$ is the magnitude
of the electric field and $m$ is the electron mass.
$a_0$ is sometimes called the strength parameter or
the normalized vector potential, and it can be shown to be both
Lorentz- and gauge-invariant~\citep{heinzl.oc.2009}. The solution
to the equations of motion, where the force is given by the Lorentz
force only, can be found in many textbooks (see \citet{Gibbon}
for example), so we will only summarize it here.

The electromagnetic field tensor for the wave is
$e F_{\mu\nu} = m a_0 \sum_i f_i'(\phi) (k_\mu \varepsilon^i_\nu - k_\nu \varepsilon^i_\mu)$,
where $k$ is the wavevector,
primes denote differentiation with respect to phase $\phi = k.x$,
and the $\varepsilon_{1,2}$ are constant polarization vectors that
satisfy $\varepsilon_i^2 = -1$ and $k.\varepsilon_i = 0$.
Then the four-momentum of the electron $p$ may be written
in terms of the potential $e A_\mu = m a_0 \sum_i f_i(\phi) \varepsilon^i_\mu$:
	\begin{equation}
	p^\mu(\phi) =
		p_0^\mu + e A^\mu
		- \left( \frac{e A.p_0}{k.p_0} + \frac{e^2 A^2}{2 k.p_0} \right) k^\mu.
	\label{eq:WaveMomentum}
	\end{equation}
Translational symmetry guarantees that $k.p = k.p_0$.
The electron trajectory $x^\mu(\phi) = \int (p^\mu / k.p) \,\rmd\phi$.

Let us say that the electron initially counterpropagates into
a circularly polarized, monochromatic wave,
with velocity $\beta_0$ and Lorentz factor $\gamma_0$.
The electron is accelerated by the wave in the longitudinal
direction, parallel to its wavevector, reaching a steady drift
velocity of $\beta_d$. Transforming to the electron's average rest
frame (ARF), as shown in \cref{fig:LaserSynch}, we find that the electron
executes circular motion with Lorentz factor
$\gamma' = (1 + a_0^2)^{1/2}$, velocity $\beta' = a_0 (1 + a_0^2)^{-1/2}$
and radius $r' = a_0 / [\gamma_0 (1 + \beta_0) \omega_0]$.
That $\gamma'$ is constant tells us that there is a phase shift
of $\pi/2$ between the rotation of the velocity and electric field
vectors $\vec{v}$ and $\vec{E}$, so $\vec{v}\cdot\vec{E} = 0$
and the external field does no work on the charge.

The instantaneous acceleration of the charge is non-zero so the
electron emits radiation while describing this orbit.
We can use classical synchrotron theory~\citep{SokolovTernov}
to calculate the energy radiated in a single cycle
$\Energy_\text{rad}$, as a fraction $f$ of the electron energy in
the ARF $\gamma' m$, with the result $f = \Energy_\text{rad}/(\gamma' m)
= 4\pi R_c/3$. The magnitude of the radiation losses is controlled
by the invariant \emph{classical radiation reaction parameter}%
~\citep{dipiazza.prl.2010}
	\begin{align}
	R_c &\equiv
		\frac{\alpha a_0^2 \gamma_0 (1 + \beta_0) \omega_0}{m}
	\label{eq:Rc}
	\\
	&\simeq 0.13
		\left(\frac{\Energy_0}{500~\text{MeV}}\right)
		\left(\frac{I_0}{10^{22}~\text{Wcm}^{-2}}\right)
		\left(\frac{\lambda}{\micron}\right).
	\end{align}
Here $\Energy_0$ is the initial energy of the electron,
$I_0 = E_0^2$ the laser intensity and $\lambda = 2\pi/\omega_0$
its wavelength.

If we define `significant' radiation damping to be an energy loss
of approximately 10\% per period~\citep{thomas.prx.2012}, we find
the threshold to be $R_c \gtrsim 0.024$, or $a_0 \gamma_0^{1/2}
\gtrsim 7\times10^2$ for a laser with a wavelength of $0.8~\micron$.
At this point the force on the electron due to radiative losses
must be included in the equations of motion. We can see this directly
by comparing the magnitudes of the radiation reaction and Lorentz
forces. Estimating the former as
$F_\text{rad} = \Energy_\text{rad} / (2\pi r')$
and the Lorentz force as $F_\text{ext} = \gamma' m / r'$,
we have that $F_\text{rad} / F_\text{ext} \simeq 2 R_c / 3$.
For $R_c \gtrsim 1$ we enter the \emph{radiation-dominated
regime}~\citep{bulanov.ppr.2004,koga.pop.2005,hadad.prd.2010}.

We will discuss how the recoil due to radiation emission is
included in classical electrodynamics in \cref{sec:RRforce}.
Before doing so, let us also consider the spectral characteristics
of the radiation emitted by the accelerated electron. In principle
the periodicity of the motion, and its infinite duration,
means that the frequency spectrum is made up of harmonics
of the ARF cyclotron frequency. However, recall that at large
$\gamma' \simeq a_0$, relativistic beaming means that most of
the radiation is emitted in the forward direction into a cone
with half-angle $1/\gamma'$. The length of the overlap between
the electron trajectory and this cone defines the \emph{formation
length} $l_f$, which is the characteristic distance over which
radiation is emitted~\citep{tm.jetp.1953,klein.rmp.1999}.
A straightforward geometrical calculation
gives the ratio between $l_f$ and the circumference of the orbit
$C = 2\pi r'$
	\begin{equation}
	\frac{l_f}{C} \simeq \frac{1}{2\pi a_0}.
	\label{eq:FormationLength}
	\end{equation}
The invariance of $a_0$ suggests we could have reached this
result in a covariant way; indeed, a full determination of
the size of the phase interval that contributes to emission
gives the same result, even quantum mechanically~\citep{ritus.jslr.1985}.

The smallness of the formation zone means that the spectrum is
broadband, with frequency components up to a characteristic value
$\omega' \simeq \gamma'^3 / r'$.
Comparing this characteristic frequency to the cyclotron frequency
(in the average rest frame) $\omega_c = 1/r'$ gives us a measure
of the classical nonlinearity:
	\begin{equation}
	\frac{\omega'}{\omega_c} \simeq a_0^3.
	\end{equation}
At $a_0 \gg 1$, the radiation is made up of very high harmonics
and is therefore well-separated from the background.
The ratio between the frequency $\omega'$
and the electron energy in the ARF $\chi = \omega' / (\gamma' m)$
is another useful invariant parameter
	\begin{align}
	\chi &\equiv \frac{a_0 \gamma_0 (1 + \beta_0) \omega_0}{m}
	\label{eq:Chi}
	\\
	&\simeq 0.29
		\left(\frac{\Energy_0}{500~\text{MeV}}\right)
		\left(\frac{I_0}{10^{22}~\text{Wcm}^{-2}}\right)^{1/2}.
	\end{align}
Restoring factors of $\hbar$ and $c$ we can show that $\chi
\propto \hbar$, unlike $R_c$. It therefore parametrizes the
importance of quantum effects on radiation reaction~\citep{ritus.jslr.1985,erber.rmp.1966},
as can be seen by the fact that if $\chi \sim 1$, an individual
photon of the radiation can carry off a substantial fraction
of the electron's energy.
By setting $\gamma_0 = 1$ in \cref{eq:Chi}, we can show that
$\chi$ is equal to the ratio of the electric field in the
instantaneous rest frame of the electron to the so-called
\emph{critical field} of QED~\citep{sauter.zp.1931,heisenberg.zp.1936}
	\begin{equation}
	\Ecrit \equiv \frac{m^2}{e} = 1.326\times10^{18}~\text{Vm}^{-1},
	\label{eq:CriticalField}
	\end{equation}
which famously marks the threshold for nonperturbative
electron-positron pair creation from the vacuum~\citep{schwinger.pr.1951}.

	\begin{figure}
	\centering
	\includegraphics[width=0.7\linewidth]{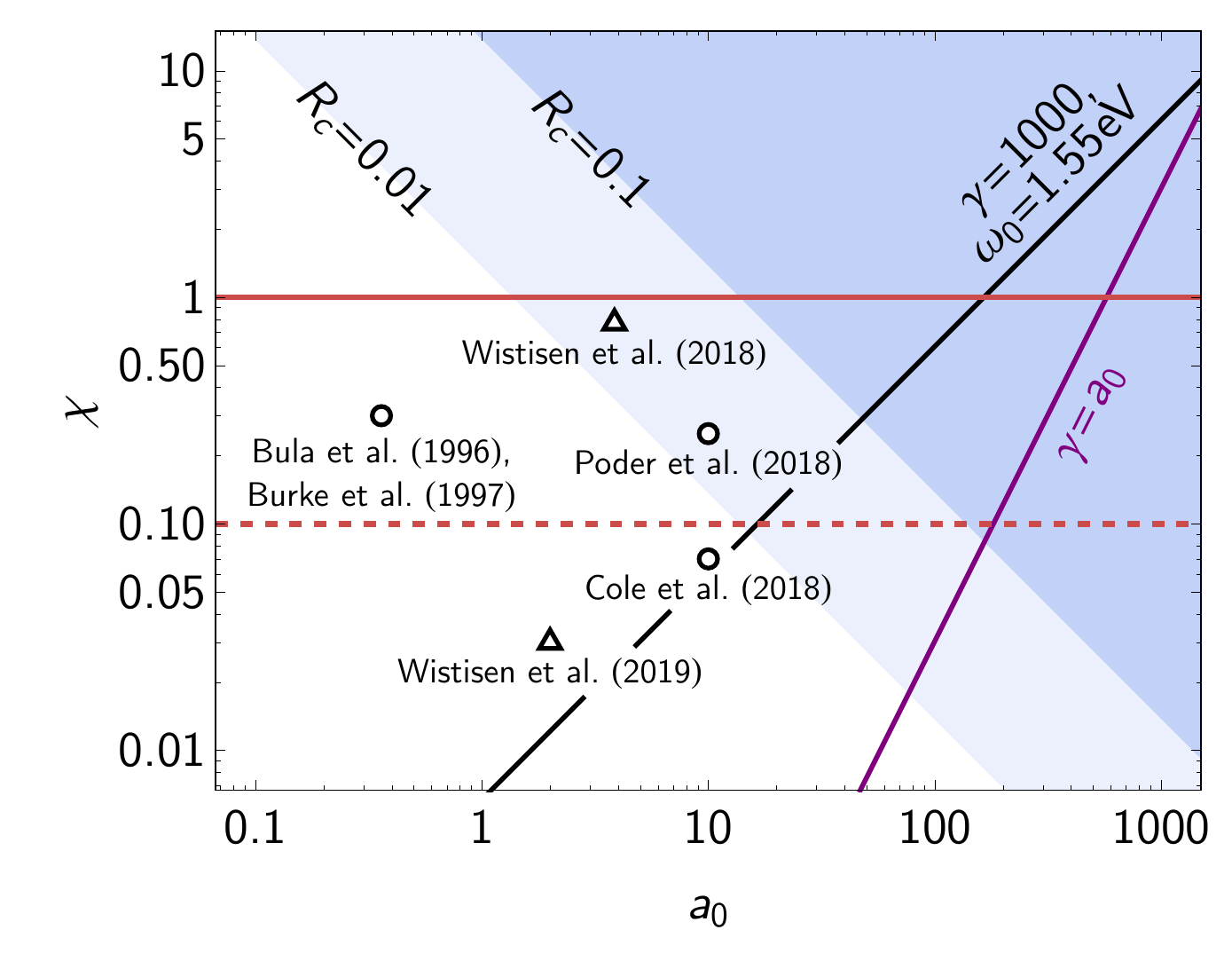}
	\caption{The importance, and type, of radiation reaction
		effects can be parametrized by
		$a_0$, the normalized intensity of the laser field or
		classical nonlinearity parameter,
		and $\chi$, the quantum nonlinearity parameter.
		Classical radiation damping becomes strong when
		$R_c = \alpha a_0 \chi > 0.01$ (light blue) and
		dominates when $R_c > 0.1$ (darker blue).
		Quantum corrections to the spectrum become necessary
		when $\chi > 0.1$. Electron-positron pair creation
		and QED cascades are important for $\chi > 1$.
		Experiments that have explored quantum effects with
		intense lasers are shown by open circles~%
		\citep{bula.prl.1996,burke.prl.1997,cole.prx.2018,%
		poder.prx.2018}.
		Two recent experiments with lepton beams and
		aligned crystals are shown by
		triangles~\citep{wistisen.ncomms.2018,wistisen.prr.2019};
		here the perpendicular component of the lepton momentum
		$p_\perp$ is used to define an equivalent classical
		nonlinearity parameter $a_0 \simeq p_\perp / m$.}
	\label{fig:a0-chi}
	\end{figure}
	
The two parameters $R_c$ and $\chi$ allow us to characterize the
importance of classical and quantum radiation reaction respectively.
We show these as functions of $a_0$ and $\chi$, the classical
and quantum nonlinearity parameters, in \cref{fig:a0-chi}.
It is evident that, as $a_0$ increases, it requires less and
less electron energy to enter the radiation-dominated regime.
Indeed, if the acceleration is provided entirely by the laser
so that $\gamma \simeq a_0$, radiation reaction becomes dominant
at about the same $a_0$ that quantum effects become important,
assuming that $\omega_0$ corresponds to a wavelength of $0.8~\micron$.
However, for $a_0 \lesssim 50$ as is accessible with existing
lasers~\citep{bahk.ol.2004,sung.ol.2017,kiriyama.ol.2018}, it is not possible to probe
radiation reaction via direct illumination of a plasma.
Instead, the experiments illustrated in \cref{fig:a0-chi} have
used pre-accelerated electrons to explore the strong-field regime,
thereby boosting both $R_c$ and $\chi$. (Note that, as $R_c$
is defined on a per-cycle basis, it would be possible for classical
radiation reaction effects to be large in long laser pulses
while remaining below the threshold for quantum effects.)
The next generation of laser facilities will reach $a_0$ in
excess of 100, perhaps even $1000$~\citep{papadopoulos.hpl.2016,weber.mre.2017,gales.rpp.2018}.
The plasma dynamics explored in such experiments will be
strongly affected by radiation reaction and quantum effects.

\section{Theory of radiation reaction}

\subsection{Classical radiation reaction}
\label{sec:RRforce}

In classical electrodynamics, radiation reaction is the response
of a charged particle to the field of its own radiation~\citep{lorentz.1904,Abraham}.
The first equation of motion to include both the external and self-induced
electromagnetic forces in a manifestly covariant and self-consistent
way was obtained by \citet{dirac.prsa.1938}. This solution
starts from the coupled Maxwell's and Lorentz equations and features
a mass renormalization that is needed to eliminate divergences
associated with a point-like charge~\citep{erber.fphys.1961,teitelboim}.
The result is generally referred to as the Lorentz-Abraham-Dirac (LAD)
equation. For an electron with four-velocity $u$, charge $-e$ and
mass $m$ it reads
	\begin{equation}
	\frac{\rmd u^\mu}{\rmd \tau} =
		-\frac{e}{m} F^{\mu\nu} u_\nu
		+ \frac{e^2}{6\pi m}
			\left(
				\frac{\rmd^2 u^\mu}{\rmd \tau^2}
				+ u^\mu \frac{\rmd u_\nu}{\rmd \tau} \frac{\rmd u^\nu}{\rmd \tau}
			\right)
	\label{eq:LAD}
	\end{equation}
where $\tau$ is the proper time. Here $F_{\mu\nu}$ is the field tensor
for the externally applied electromagnetic field, so it is the second
term that accounts for the self-force. Although the LAD equation is
an exact solution of the Maxwell-Lorentz system, using it directly
turns out to be problematic. The momentum derivative
$\frac{\rmd^2 u^\mu}{\rmd\tau^2}$ in the RR term leads
to so-called \emph{runaway} solutions, in which the electron energy
increases exponentially in the absence of external fields, and to
\emph{pre-acceleration}, in which the momentum changes in advance
of a change in the applied field~\citep{Spohn,Yaghjian,Rohrlich}.
These issues have prompted searches for alternative classical
theories of radiation reaction%
~\citep{mo.papas,bonnor,eliezer,ford.oconnell,sokolov.jetp.2009}
that have more satisfactory properties (see the review by
\cite{burton.cp.2014} for details).

The most widely used classical theory is that proposed by
\cite{landau.lifshitz}. They realized that if the second
(RR) term in \cref{eq:LAD} were much smaller than the first in the
instantaneous rest frame of the charge, it would be possible to
reduce the order of the LAD equation by substituting
$\frac{\rmd u}{\rmd\tau} \to \frac{e}{m} F^{\mu\nu} u_\nu$ in the RR term.
The result, called the Landau-Lifshitz equation, is first-order in
the electron momentum and free from the pathological solutions of
the LAD equation~\citep{burton.cp.2014}:
	\begin{equation}
	\frac{\rmd u^\mu}{\rmd \tau} =
		- \frac{e}{m} F^{\mu\nu} u_\nu
		+ \frac{e^4}{6\pi m}
			\left[
				-\frac{m}{e}(\partial_\alpha F^{\mu\nu}) u_\nu u^\alpha
	\right. \\ \left.
				+ F^{\mu\nu} F_{\nu\alpha} u^\alpha
				+ (F^{\nu \alpha} u_\alpha)^2 u^\mu
			\right].
	\label{eq:LandauLifshitz}
	\end{equation}
The following two conditions for the characteristic length scale $L$ over
which the field varies and its magnitude $E$ must be fulfilled in
the instantaneous rest frame for the order reduction procedure to be valid:
$L \gg \lambda_C$ and $E \ll \Ecrit/\alpha$,
where $\lambda_C = 1/m$ is the Compton length.
Note that both of these are automatically fulfilled in the realm of
classical electrodynamics~\citep{dipiazza.rmp.2012}, as quantum effects
can only be neglected when $L \gg \lambda_C$ and $E \ll \Ecrit$.
The former condition ensures that the electron wavefunction is
well-localized and the latter means recoil at the level of the
individual photon is negligible~\citep{dipiazza.rmp.2012}.
One reason to favour the Landau-Lifshitz equation is that all
physical solutions of the LAD equation are solutions of the
Landau-Lifshitz equation~\citep{spohn.euro.2000}.

Once the trajectories are determined, the self-consistent radiation
is obtained from the Li{\'e}nard-Wiechert potentials,
which give the electric and magnetic fields of a charge in
arbitrary motion~\citep{Jackson}.
The spectral intensity of the radiation from an ensemble of $N_e$
electrons, the energy radiated per unit frequency $\omega$ and solid angle
$\Omega$, is given in the far field by
	\begin{equation}
	\frac{\rmd^2 \Energy}{\rmd \omega \rmd \Omega} =
		\frac{\alpha \omega^2}{4 \pi^2}
		\left| \sum_{k=1}^{N_e} \int_{-\infty}^\infty 
			\vec{n} \times (\vec{n} \times \vec{v}_k)
			e^{i \omega (t - \vec{n}\cdot\vec{r}_k)}
		\,\rmd t \right|^2
	\label{eq:SpectralIntensity}
	\end{equation}
where $\vec{n}$ the observation direction, and $\vec{r}_k$
and $\vec{v}_k$ are the position and velocity of the $k$th
particle at time $t$~\citep{Jackson}.

\subsubsection{In plane electromagnetic waves}

Among the other useful properties of \cref{eq:LandauLifshitz} is that
it can be solved exactly if the external field is a plane electromagnetic
wave~\citep{dipiazza.lmp.2008}.
Taking this field to be
$e F_{\mu\nu} = m a_0 \sum_i f_i'(\phi) (k_\mu \varepsilon^i_\nu - k_\nu \varepsilon^i_\mu)$,
using the same definitions as in \cref{sec:Parameters},
\cref{eq:LandauLifshitz} is most conveniently expressed in terms of
the \emph{lightfront} momentum $\um \equiv k.p/(m \omega_0)$,
scaled perpendicular momenta $\tilde{u}_{x,y} \equiv u_{x,y}/\um$,
and phase $\phi$:
	\begin{equation}
	\frac{\rmd\um}{\rmd \phi} =
		-\frac{2\alpha a_0^2 \omega_0}{3 m}
		[ f'_1(\phi)^2 + f'_2(\phi)^2 ] {\um}^2,
	\label{eq:LLminus}
	\end{equation}
and
	\begin{equation}
	\frac{\rmd \tilde{u}_i}{\rmd \phi} =
		\frac{a_0 f_i'(\phi)}{u^-} + \frac{2 \alpha a_0 \omega_0 f_i''(\phi)}{3 m}.
 	\label{eq:LLperp}
	\end{equation}
The remaining component $\up$ is determined by the mass-shell
condition $\um\up - u_x^2 - u_y^2 = 1$ and the position by integration of
$\omega_0 x^\mu(\phi) = \int_{-\infty}^\phi (u^\mu/u^-) \,\rmd\phi$.
\Cref{eq:LLminus} admits the solution
	\begin{equation}
	\um =
		\frac{\um_0}{1 + \frac{2}{3} R_c I(\phi)},
	\label{eq:LLminusSol}
	\end{equation}
where $\um_0$ is the initial lightfront momentum, the classical radiation
reaction parameter $R_c = a_0^2 \um_0 \omega_0 / m$ as in \cref{eq:Rc},
and $I(\phi) = \int_{-\infty}^\phi [f_1'(\psi)^2 + f_2'(\psi)^2]\,\rmd\psi$.
The choice of notation here reflects the fact that $f'(\phi)$ is proportional
to the electric field and so $I(\phi)$ is like an integrated energy flux.
We use \cref{eq:LLminusSol} to solve \cref{eq:LLperp}, obtaining
$\tilde{u}_i(\phi)$ and then
	\begin{equation}
	u_i =
		\frac{1}{1 + \frac{2}{3} R_c I(\phi)}
		\left[
			u_{i,0} + a_0 f_i(\phi)
			+ \frac{2 R_c}{3} H(\phi) 
			+ \frac{2 R_c}{3 a_0} f_i'(\phi)
		\right],
	\label{eq:LLperpSol}
	\end{equation}
where $u_{i,0}$ is the initial value of the perpendicular momentum component $i$
and $H_i(\phi) = \int_{-\infty}^\phi\! f_i'(\psi) I(\psi) \,\rmd\psi$.
The electron trajectory in the absence of radiation reaction is obtained by setting
$\alpha = 0$, in which case we recover \cref{eq:WaveMomentum} as expected.
Note that the lightfront momentum $u^-$ is no longer conserved, once
radiation reaction is taken into account~\citep{harvey.prd.2011}.

In \cref{sec:Parameters} we estimated that the electron would radiate
in a single cycle a fraction $4\pi R_c/3$ of its total energy.
Using our analytical result \cref{eq:LLminusSol} and assuming
$\gamma \gg 1$ so that $\um \simeq 2\gamma$, we can show this fraction is actually
$\Energy_\text{rad}/(\gamma_0 m) = (4\pi R_c/3) / (1 + 4\pi R_c /3)$.
Here the denominator represents radiation-reaction corrections to
the energy loss, guaranteeing that $\Energy_\text{rad}/(\gamma_0 m) < 1$.
With these corrections, the energy emitted, according to
the Larmor formula, is equal to the energy lost, according to the
Landau-Lifshitz equation (see Appendix A of \citet{dipiazza.plb.2018}
for a direct calculation of momentum conservation).

The emission spectrum \cref{eq:SpectralIntensity} may also be
expressed in terms of an integral over phase.
The number of photons scattered per unit (scaled) frequency $s = \omega/\omega_0$
and solid angle is~\citep{esarey.pre.1993,hartemann.prl.2010}
	\begin{equation}
	\frac{\rmd^2 N_\gamma}{\rmd s \rmd\Omega} =
		\frac{\alpha s}{4 \pi^2}
		\left|
			\sum_{k=1}^{N_e} \int_{-\infty}^{\infty} \!
				\frac{\varepsilon'.u_k \exp(-i s n.\xi_k)}{(\um_k)^2}
			\, \rmd \phi
		\right|^2
	\label{eq:PhotonSpectrum}
	\end{equation}
where the scaled four-position $\xi \equiv \omega_0 x$, and $\varepsilon'$
and $n$ are the four-polarization and propagation direction
of the scattered photon.
Given these relations and the analytically determined trajectory, we can
numerically evaluate the number of photons scattered to given frequency
and polar angle by integrating \cref{eq:PhotonSpectrum}, summed
over polarizations, over all azimuthal angles $0 \leq \varphi < 2\pi$.

\subsection{Quantum corrections: suppression and stochasticity}
\label{sec:QuantumCorrections}

We showed in \cref{fig:a0-chi} that in many scenarios of
interest, reaching the regime where radiation reaction becomes
important automatically makes quantum effects important as well.
This raises the question: what is the quantum picture of radiation
reaction? Let us revisit the example we studied classically in
\cref{sec:Parameters}, that of an electron emitting radiation
under acceleration by a strong electromagnetic wave. One might
instinctively liken this scenario to inverse Compton scattering,
as energy and momentum are automatically conserved when the electron
absorbs a photon (or photons) from the plane wave and emits
another, higher energy photon. However, the recoil is proportional
to $\hbar$ and vanishes in the classical limit; we would then
recover Thomson scattering rather than radiation reaction.

The solution is that, in the regime $a_0 \gg 1$ and $\chi \lesssim 1$,
quantum radiation reaction can be identified with the recoil
on the electron due its emission of multiple, incoherent
photons~\citep{dipiazza.prl.2010}. These conditions express the
following: $a_0 \gg 1$ means that the formation length is
much smaller than the wavelength of the external field,
by \cref{eq:FormationLength}, so the coherent contribution is
suppressed; and $\chi \lesssim 1$ means pair creation can be
neglected. The latter is important because QED is inherently
a many-body theory and it is possible for the final state
to contain many more electrons than the initial state.
As the number of photons $N_\gamma \propto \alpha \propto 1/\hbar$
and the momentum change of the electron $\propto \hbar$ for each
photon, we have that the total momentum change $\propto \hbar^0$
and therefore a classical limit exists~\citep{dipiazza.rmp.2012}.
This suggests that one way to determine the `correct' theory
of classical radiation reaction is to start with a QED result and
take the limit $\hbar \to 0$. This has been accomplished for
both the momentum change~\citep{krivitski.spu.1991,ilderton.plb.2013}
and the position~\citep{ilderton.prd.2013}. In particular,
\citet{ilderton.prd.2013} were able to
show that, to first order in $\alpha$, only the LAD, Landau-Lifshitz
and Eliezer-Ford-O'Connell formulations of
radiation reaction were consistent with QED.

In both the classical and quantum regimes, the force of radiation
reaction is directed antiparallel to the electron's instantaneous
momentum, and its magnitude depends on the parameter $\chi$.
We defined this earlier for the particular case of an electron in
an electromagnetic wave [see \cref{eq:Chi}].
In a general electromagnetic field $F_{\mu\nu}$,
	\begin{equation}
	\chi = \frac{\sqrt{-(F_{\mu\nu} p^\nu)^2}}{m \Ecrit} =
		\frac{\gamma}{\Ecrit}
		\sqrt{(\vec{E} + \vec{v}\times\vec{B})^2 - (\vec{v}\cdot\vec{E})^2},
	\label{eq:ChiGeneral}
	\end{equation}
where $p = \gamma m (1, \vec{v})$ is the electron four-momentum.
$\chi$ depends on the instantaneous transverse acceleration induced
by the external field: in a plane EM wave, where $\vec{E}$ and $\vec{B}$
have the same magnitude and are perpendicular to each other,
$\chi = \gamma \abs{\vec{E}} (1 - \cos\theta) / \Ecrit$, where $\theta$ is the angle
between the electron momentum and the laser wavevector, and it is
therefore largest in counterpropagation.
A curious consequence of \cref{eq:ChiGeneral} is the existence of a \emph{radiation-free
direction}: no matter the configuration of $\vec{E}$ and $\vec{B}$, there
exists a particular $\vec{v}$ that makes $\chi$ vanish~\citep{gonoskov.pop.2018}.
Electrons in extremely strong fields tend to align themselves
with this direction, any transverse momentum they have being rapidly
radiated away~\citep{gonoskov.pop.2018}.
As this direction is determined purely by the fields, the
self-consistent evolutions of particles and fields is determined
by hydrodynamic equations~\citep{samsonov.pra.2018}.

\begin{figure}
	\centering
	\includegraphics[width=\linewidth]{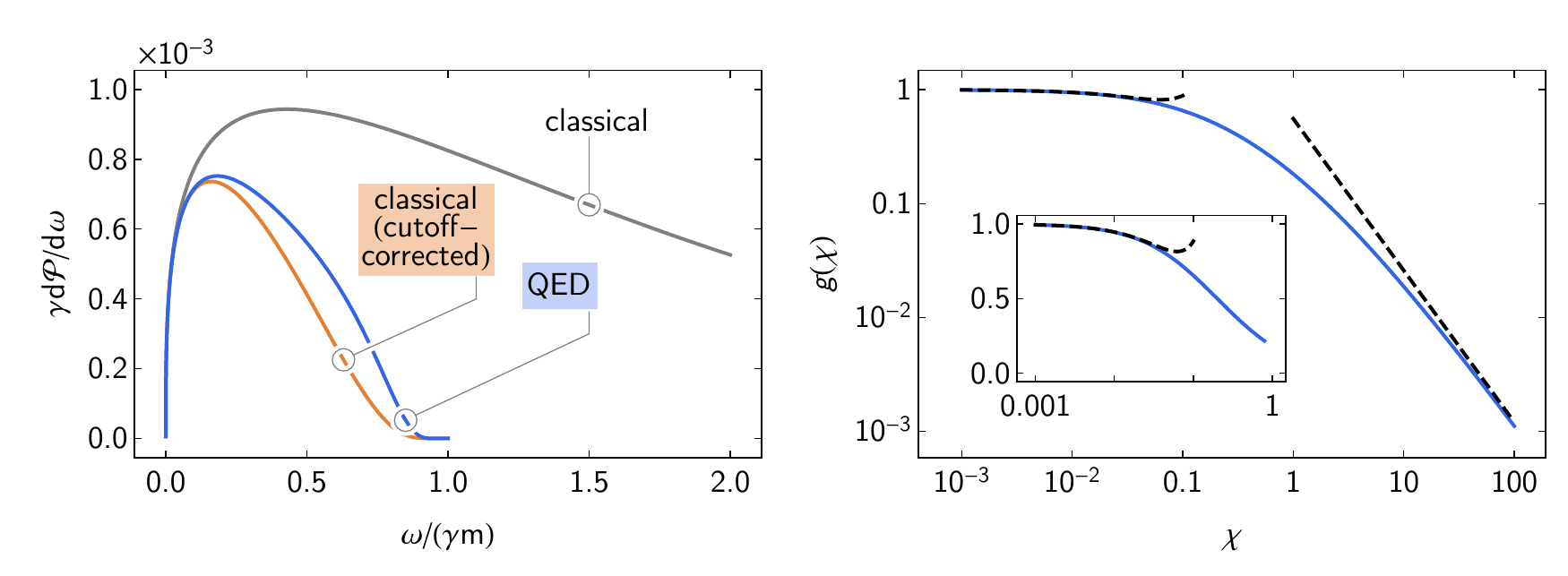}
	\caption{%
		(left) Quantum corrections to the emission spectrum
		$\rmd \Power / \rmd \omega$ at $\chi = 1$:
		the classical [\cref{eq:ClassicalSpectrum}]
		and quantum-corrected spectra [\cref{eq:QuantumSpectrum}].
		(right) These corrections cause the total radiated power
		to be reduced by a factor $g(\chi)$: the full result (blue)
		and limiting expressions (black, dashed).}
	\label{fig:QuantumCorrections}
	\end{figure}

The larger the value of $\chi$, the greater the differences between the quantum
and classical predictions of radiation emission. Classically there is no upper
limit on the frequency spectrum, whereas in the quantum
theory there appears a cutoff that guarantees $\omega < \gamma m$.
Besides this cutoff, spin-flip transitions enhance the spectrum
at high energy~\citep{uggerhoj.rmp.2005}.
Let us work in the synchrotron limit, wherein the field may be considered
constant over the formation length (i.e. $l_f \ll \lambda$, using \cref{eq:FormationLength}).
The classical emission spectrum, the energy radiated per unit frequency
$\omega = x \gamma m$ and time by an electron with quantum parameter
$\chi$ and Lorentz factor $\gamma$, is
	\begin{align}
	\frac{\rmd \Power_\text{cl}}{\rmd \omega} &=
		\frac{\alpha \omega}{\sqrt{3}\pi\gamma^2}
		\left[
			2 K_{2/3}(\xi) - \int_\xi^\infty \!K_{1/3}(y) \,\rmd y
		\right],
	&
	\xi &= \frac{2 x}{3\chi}.
	\label{eq:ClassicalSpectrum}
	\end{align}
Two quantum corrections emerge when $\chi$ is no longer much smaller than
one: the non-negligible recoil of an individual photon means
that the spectrum has a cutoff at $x = 1$; and the spin contribution
to the radiation must be included.
The former can be included directly by modifying
$\xi = 2x / (3\chi) \to 2x / [3\chi(1-x)]$ in \cref{eq:ClassicalSpectrum},
which yields the spectrum of a \emph{spinless} electron (shown in
orange in \cref{fig:QuantumCorrections}).
A neat exposition of this simple substitution is given by \citet{lindhard.pra.1991}
in terms of the correspondence principle (see also \citet{sorensen.nimb.1996}).
Then when the spin contribution is added, we obtain the full QED result%
~\citep{erber.rmp.1966,ritus.jslr.1985,Baier}
	\begin{align}
	\frac{\rmd \Power_\text{q}}{\rmd \omega} &=
		\frac{\alpha \omega}{\sqrt{3}\pi\gamma^2}
		\left[
			\left( 1 - x + \frac{1}{1 - x} \right) K_{2/3}(\xi) - \int_\xi^\infty \!K_{1/3}(y) \,\rmd y
		\right],
	&
	\xi &= \frac{2 x}{3\chi (1 - x)},
	\label{eq:QuantumSpectrum}
	\end{align}
where we quote the spin-averaged and polarization-summed result.
This is shown in blue in \cref{fig:QuantumCorrections}.
The number spectrum $\frac{\rmd N_\gamma}{\rmd \omega} =
\omega^{-1} \frac{\rmd \Power_\text{q}}{\rmd \omega}(\chi,\gamma)$
has an integrable singularity $\propto \omega^{-2/3}$ in the
limit $\omega \to 0$. The total number of photons
$N_\gamma = \int \frac{\rmd N_\gamma}{\rmd \omega} \rmd\omega$ is finite.

The combined effect of these corrections is to reduce the instantaneous
power radiated by an electron. This reduction is quantified by the factor
$g(\chi) = \Power_\text{q} / \Power_\text{cl}$, which takes
the form~\citep{erber.rmp.1966,SokolovTernov}
	\begin{align}
	g(\chi) &=
		\frac{9\sqrt{3}}{8\pi}
		\int_0^\infty
			\left[
			\frac{2 u^2 K_{5/3}(u)}{(2 + 3 \chi u)^2}
			+ \frac{36 \chi^2 u^3 K_{2/3}(u)}{(2 + 3 \chi u)^4}
			\right]
		\rmd u
	\label{eq:Gaunt}
	\\
	&=	\begin{cases}
		1 - \frac{55\sqrt{3}}{16}\chi + 48\chi^2 & \chi \ll 1 \\
		\frac{16 \Gamma(2/3)}{3^{1/3} 27} \chi^{-4/3} & \chi \gg 1
		\end{cases}
	\label{eq:GauntLimits}
	\end{align}
where $K$ is a modified Bessel function of the second kind
and $\Gamma(2/3) \simeq 1.354$.
The limiting expressions given in \cref{eq:GauntLimits} are within
5\% of the full result for $\chi < 0.05$ and $\chi > 200$ respectively.
A simple analytical approximation to \cref{eq:Gaunt} that is
accurate to 2\% for arbitrary $\chi$ is $g(\chi) \simeq
[1 + 4.8(1+\chi)\ln(1+1.7\chi) + 2.44\chi^2]^{-2/3}$~\citep{Baier}.
The changes to the classical radiation spectrum and the magnitude
of $g(\chi)$ are shown in \cref{fig:QuantumCorrections}.
Note that the total power $\Power_\text{q} = 2\alpha m^2 \chi^2 g(\chi)/3$
always increases with increasing $\chi$.
$g(\chi)$ is sometimes referred to as the
`Gaunt factor'~\citep{ridgers.jpp.2017}, as it is a multiplicative
(quantum) correction to a classical result, first derived in
the context of absorption~\citep{gaunt.pt.1930}.

\Cref{fig:QuantumCorrections} shows that the radiated power at
$\chi \sim 1$ is less than $20\%$ of its classically predicted value.
While this suppression does have a marked effect on the particle
dynamics, it is not the only quantum effect. As is discussed in
\cref{sec:Parameters}, $\chi$ is the ratio between the energies
of the typical photon and the emitting electron. When this approaches
unity, even a single emission can carry off a large fraction of
the electron energy, and the concept of a continuously radiating
particle breaks down.
Instead, electrons lose energy \emph{probabilistically}, in discrete
portions. The importance of this discreteness may be estimated
by comparing the typical time interval between emissions,
$\Delta t = \avg{\omega}/\Power$, with the timescale of the
laser field $1/\omega_0$~\citep{gonoskov.pre.2015}.
\Cref{eq:QuantumSpectrum} yields for the average photon energy
$\avg{\omega} \simeq 0.429\, \chi \gamma m$
for $\chi \ll 1$ and $0.25\, \gamma m$ for $\chi \gg 1$; the
radiated power $\Power = 2\alpha m^2 \chi^2 g(\chi)/3$. We find
	\begin{equation}
	\omega_0 \Delta t \simeq
		\begin{cases}
		44\, a_0^{-1} & \chi \ll 1 \\
		58\, [\gamma \omega_0 / (a_0 m)]^{1/3} & \chi \gg 1
		\end{cases}.
	\end{equation}
We expect stochastic effects to be at their most significant when
$\omega_0 \Delta t \gtrsim 1$, which implies that the total number of
emissions in an interaction is relatively small but $\chi$ is large.

	\begin{figure}
	\centering
	\includegraphics[width=0.3\linewidth]{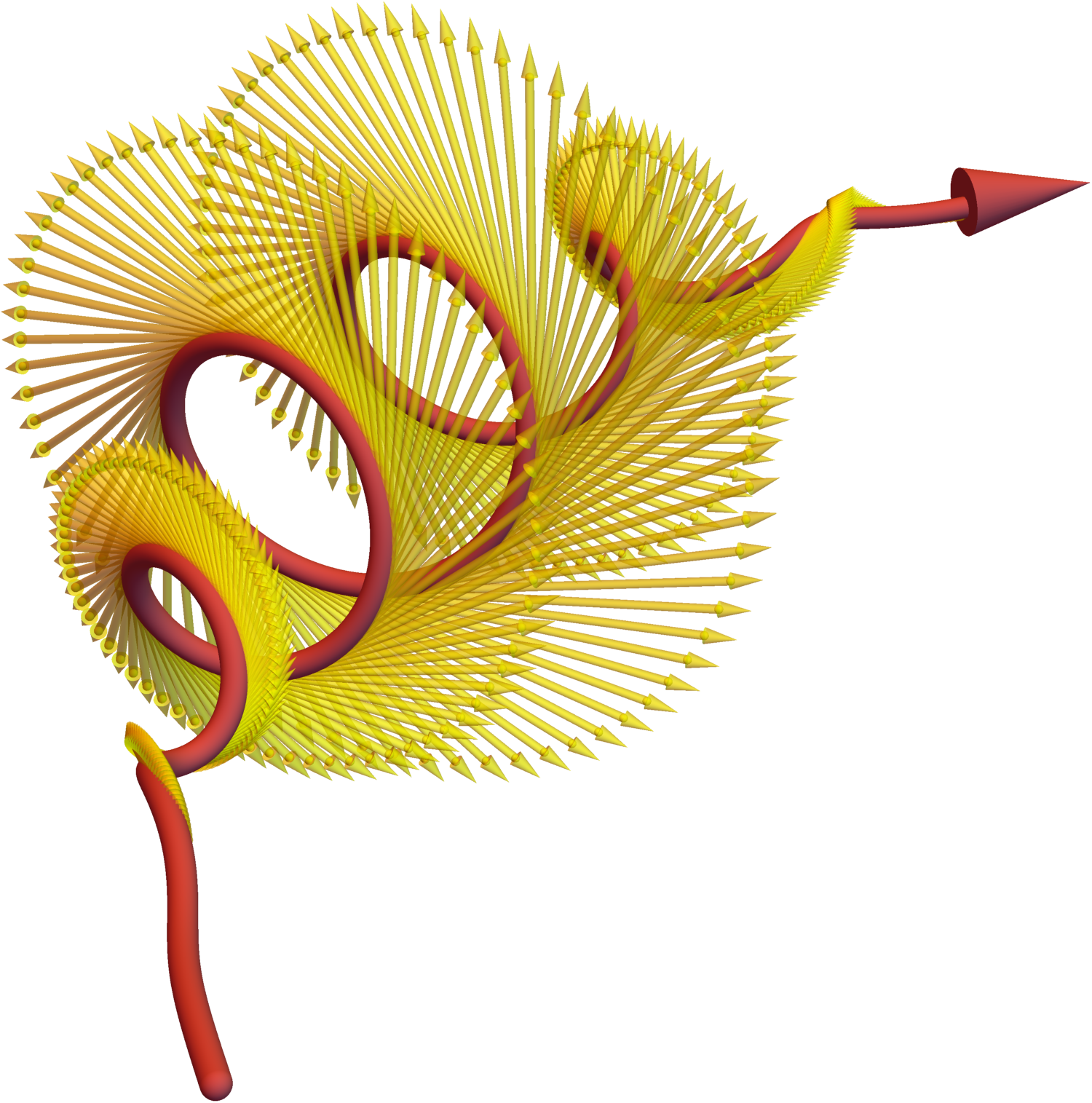}
	\includegraphics[width=0.3\linewidth]{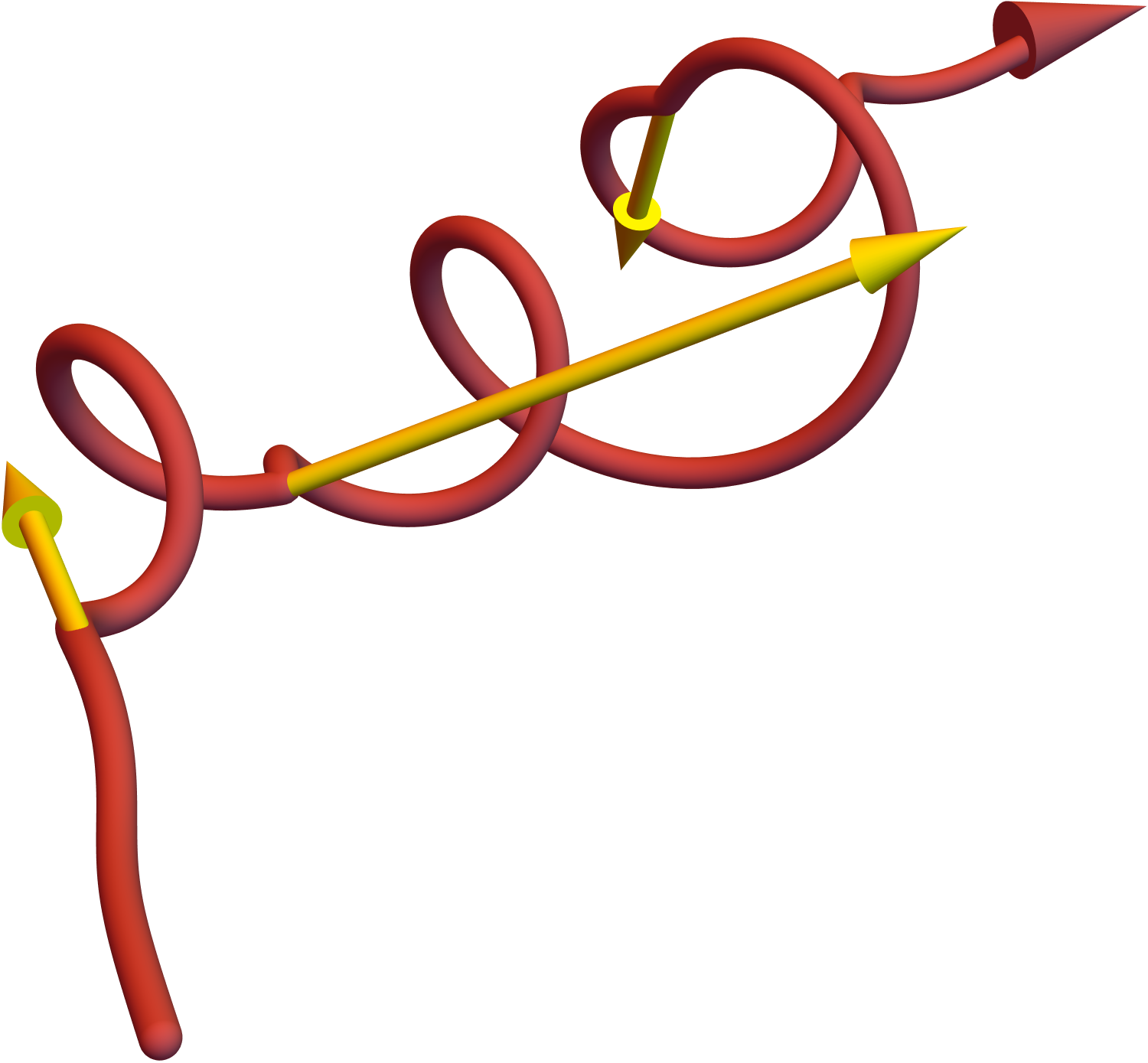}
	\caption{
		In the classical picture, radiation reaction is a continuous drag
		force that arises from the emission of very many photons that individually
		have vanishingly low energies (left).
		In the quantum regime, however, the electrons emits a finite number of
		photons, any or all of which can exert a significant recoil on
		the electron.
		The probabilistic nature of emission leads to radically altered
		electron dynamics, with implications for laser-matter interactions
		beyond the current intensity frontier.
		From \citet{blackburn.thesis.2015}.
		}
	\label{fig:RR}
	\end{figure}

A description of how stochastic energy losses can be modelled
follows in \cref{sec:ModellingQuantum}. For now, it suffices to
interpret \cref{eq:QuantumSpectrum} as the (energy-weighted)
probability distribution of the photons emitted at a particular
instant of time. Even though two electrons may have the same
$\chi$ and $\gamma$, they can emit photons of different energies
(or none at all) and thereby experience different recoils.
Contrast this with the classical picture, in which the continuous
energy loss is driven by the emission of many photons that
individually have vanishingly small energies (see \cref{fig:RR}).

Consider, for example, the interaction of a beam of electrons with
a plane electromagnetic wave, where the Lorentz factors of the
electrons are distributed $\gamma \sim \frac{\rmd N_e}{\rmd \gamma}$.
The distribution is characterized by a mean $\mu \equiv \avg{\gamma}$
and variance $\sigma^2 \equiv \avg{\gamma^2} - \mu^2$.
Under classical radiation reaction, higher energy electrons are
guaranteed to radiate more than their lower energy counterparts
($\Power \propto \gamma^2$), with the result that both the mean
and the variance of $\gamma$ decrease over the course of the
interaction~\citep{neitz.prl.2013}. This is still the case if
the radiated power is reduced by the Gaunt factor $g(\chi)$,
i.e. a `modified classical' model is assumed (see \cref{sec:ModellingClassical}),
because radiation losses remain deterministic~\citep{yoffe.njp.2015}.

Under quantum radiation reaction, radiation losses are inherently
probabilistic. While $\mu$ will still decrease (more energetic
electrons radiate more energy \emph{on average}), the width of
the distribution $\sigma^2$ can actually grow~\citep{neitz.prl.2013,%
vranic.njp.2016}. \Citet{ridgers.jpp.2017} derive the following
equations for the temporal evolution of these quantities, under
quantum radiation reaction:
	\begin{align}
	\frac{\rmd \mu}{\rmd t} &=
		- \frac{2\alpha m}{3} \avg{\chi^2 g(\chi)}
	\label{eq:MuEvolution}
	\\
	\frac{\rmd \sigma^2}{\rmd t} &=
		-\frac{4\alpha m}{3} \avg{(\gamma - \mu) \chi^2 g(\chi)}
		+ \frac{55 \alpha m}{24 \sqrt{3}} \avg{\gamma \chi^3 g_2(\chi)},
	\label{eq:VarianceEvolution}
	\end{align}
where $\avg{\cdots}$ denotes the population average and $g_2(\chi) =
\int\!\chi\,\rmd \Power_\text{q} / \int\!\chi\,\rmd \Power_\text{cl}$
is the second moment of the emission spectrum.
Only the first term of \cref{eq:VarianceEvolution} is non-zero in
the classical limit, and it is guaranteed to be negative.
The second term represents stochastic effects and is always positive.
Broadly speaking, the latter is dominant if $\chi$ is large,
the interaction is short, or the initial variance is small%
~\citep{ridgers.jpp.2017,niel.pre.2018}.
The evolution of higher order moments, such as the skewness of
the distribution, are considered in \citet{niel.pre.2018}.

A distinct consequence of stochasticity is \emph{straggling}%
~\citep{shen.prl.1972}, where an electron that radiates
less (or no) energy than expected enters regions of
phase space that would otherwise be forbidden. Unlike stochastic
broadening, which can occur in a static, homogeneous electromagnetic
field, straggling requires the field to have some non-trivial
spatiotemporal structure. If an unusually long interval passes
between emissions, an electron may be accelerated to a higher energy
or sample the fields at locations other than those along the
classical trajectory~\citep{duclous.ppcf.2011}.
In a laser pulse with a temporal envelope, for example, electrons
that traverse the intensity ramp without radiating reach larger
values of $\chi$ than would be possible under continuous radiation
reaction; this enhances high-energy photon production and
electron-positron pair creation~\citep{blackburn.prl.2014}.
If the laser duration is short enough, it is probable that the
electron passes through the pulse without emitting at all, in
so-called \emph{quenching} of radiation losses~\citep{harvey.prl.2017}.

The quantum effects we have discussed in this section emerge,
in principle, from analytical results including the emission
spectrum [\cref{eq:QuantumSpectrum}]. While further analytical progress
can be made in the quantum regime, using the theory
of \emph{strong-field QED} (see \cref{sec:ModellingQuantum}),
modelling more realistic laser--electron-beam or laser-plasma
interactions generally requires numerical simulations.
Much effort has been devoted to the development, improvement,
benchmarking and deployment of such simulation tools over
the last few years. In the following section we review
these continuing developments.

\section{Numerical modelling and simulations}

\subsection{Classical regime}
\label{sec:ModellingClassical}

A natural starting point is the modelling of classical radiation
reaction effects. In the absence of quantum corrections,
we have all the ingredients we need to formulate a self-consistent
picture of radiation emission and radiation reaction.
We showed in \cref{sec:Parameters} how using only the Lorentz force to
determine the charge's motion and therefore its emission led
to an inconsistency in energy balance.
This is remedied by using either \cref{eq:LAD} or \cref{eq:LandauLifshitz}
as the equation of motion, in which case the energy carried away in
radiation matches that which is lost by the electron.

Implementations of classical radiation reaction in plasma simulation
codes have largely favoured the Landau-Lifshitz equation (or a
high-energy approximation thereto), as it is first-order in the
momentum and the additional computational cost is not
large~\citep{tamburini.njp.2010,harvey.prd.2011,green.cpc.2015,vranic.cpc.2016}.
These codes have not only been used to study radiation reaction
effects in laser-plasma interactions~\citep{chen.ppcf.2011,%
nakamura.prl.2012,vranic.prl.2014,tamburini.pre.2014,yoffe.njp.2015,liseykina.njp.2016},
but also whether there are observable differences between models of
the same~\citep{bulanov.pre.2011,kravets.pre.2013}.
The radiation reaction force proposed by \citet{sokolov.jetp.2009}
has also been implemented in some codes~\citep{sokolov.pop.2011,capdessus.pre.2012},
but note that it is not consistent with the classical limit of
QED~\citep{ilderton.prd.2013}.
It is also possible to solve the LAD equation numerically via integration
backward in time~\citep{koga.pre.2004}.

Given data on the trajectories of an ensemble of electrons
(usually a subset of the all electrons in the simulations),
\cref{eq:SpectralIntensity} can be used to obtain the far-field
spectrum in a simulation where classical radiation reaction
effects are included~\citep{thomas.prx.2012,schlegel.njp.2012,martins.ppcf.2016}.
\Cref{eq:SpectralIntensity} is valid across the full range of
$\omega$ (\emph{pace} the quantum cutoff at $\omega = \gamma m$),
including the low-frequency region of the spectrum where collective
effects are important: $\omega < n_e^{1/3}$, where $n_e$ is the
electron number density.
This region does not, however, contribute very much to radiation reaction;
this is dominated by photons near the synchrotron critical energy
$\omega_c \gg n_e^{1/3}$.
Thus the spectrum can be divided into \emph{coherent} and
\emph{incoherent} parts, that are well separated in terms of
their energy~\citep{gonoskov.pre.2015}. In the latter region,
the order of the summation and integration in \cref{eq:SpectralIntensity}
can be exchanged, and the total spectrum determined by summing over
the single-particle spectra.

In a particle-in-cell code for example, the electromagnetic field is
defined on a grid of discrete points and advanced self-consistently
using currents that are deposited onto the same grid~\citep{dawson.rmp.1983}.
Defining the grid spacing to be $\Delta$, this scheme will directly
resolve electromagnetic radiation that has a frequency less than
the Nyquist frequency $\pi/\Delta$.\footnote{Sampling at discrete
points, i.e. with limited sampling rate, means that only a certain
range of frequency components in a given waveform can be represented.
The highest frequency is called the `Nyquist frequency'.
Modes that lie are above this are aliased to lower frequencies.}
Given appropriately high resolution, this accounts for the coherent
radiation generated by the collective dynamics of the ensemble of particles.
The recoil arising from higher frequency components, which
cannot be resolved on the grid, and in any case as a self-interaction
is neglected, is accounted for by the
radiation reaction force.

Further simplification is possible if the interference of emission
from different parts of the trajectory is negligible. As indicated in
\cref{sec:Parameters}, at high intensity $a_0 \gg 1$, the formation
length of the radiation is much smaller than the timescale of the
external field (see \cref{eq:FormationLength}).
This being the case,
rather than using \cref{eq:SpectralIntensity},
we may integrate the \emph{local} emission spectrum
\cref{eq:ClassicalSpectrum} over the particle trajectory,
assuming that, at high $\gamma$, the radiation is emitted predominantly
in direction parallel to the electron's instantaneous velocity~\citep{esarey.pre.1993,reville.apj.2010,wallin.pop.2015}.
The approach is naturally extended to account for quantum effects,
by substituting for the classical synchrotron spectrum \cref{eq:ClassicalSpectrum}
the equivalent result in QED, \cref{eq:QuantumSpectrum}.

One consequence of doing so is that the radiated power is reduced by
the factor $g(\chi)$, given in \cref{eq:Gaunt}.
This should be reflected in a reduction in the magnitude of the
radiation-reaction force.
Consequently, a straightforward, phenomenological way to model quantum radiation
reaction is to use a version of \cref{eq:LandauLifshitz}
where the second term is scaled by $g(\chi)$.
This `modified classical' model has been used in studies of
laser--electron-beam~\citep{thomas.prx.2012,blackburn.prl.2014,yoffe.njp.2015}
and laser-plasma interactions~\citep{kirk.ppcf.2009,zhang.njp.2015}
as a basis of comparison with a fully stochastic model (shortly to be introduced),
as well as in experimental data analysis~\citep{wistisen.ncomms.2018,poder.prx.2018}.
It has been shown that this approach yields the
correct equation of motion for the \emph{average} energy of
an ensemble of electrons in the quantum regime~\citep{ridgers.jpp.2017,niel.pre.2018}.
It is, however, deterministic, and therefore neglects the
stochastic effects we discussed in \cref{sec:QuantumCorrections}.

\subsection{Quantum regime: the `semiclassical' approach}
\label{sec:ModellingQuantum}

In \cref{sec:QuantumCorrections} we discussed how `quantum radiation
reaction' could be identified with the recoil arising from multiple,
incoherent emission of photons. Indeed, if $\chi \gtrsim 1$,
any or all of these photons can exert a significant momentum change
individually. \Cref{fig:a0-chi} tells us that we generally require
$a_0 \gg 1$ to enter the quantum radiation reaction regime with lasers,
which necessitates a nonperturbative approach to the theory.
This is provided by \emph{strong-field} QED, which separates the
electromagnetic field into a fixed background, treated exactly,
and a fluctuating part, treated perturbatively~\citep{furry.pr.1951};
see the reviews by \citet{ritus.jslr.1985,dipiazza.rmp.2012,heinzl.ijmp.2012}
or a tutorial overview by \citet{seipt.2017} which discusses photon
emission in particular.

Although it is the most general and accurate approach, strong-field
QED is seldom used to model experimentally relevant configurations
of laser-electron interaction~\citep{blackburn.pop.2018}.
In a scattering-matrix calculation, the object is to obtain
the probability of transition between asymptotic free states;
as such, complete information about the spatiotemporal structure
of the background field is required. Analytical results have only
been obtained in field configurations that possess high symmetry%
~\citep{heinzl.prl.2017}, e.g. plane EM waves~\citep{ritus.jslr.1985}
or static magnetic fields~\citep{erber.rmp.1966}.
The assumption that the background is fixed also means that
back-reaction effects are neglected, even though it is expected that
QED cascades will cause significant depletion of energy from those
background fields~\citep{bell.prl.2008,fedotov.prl.2010,bulanov.prl.2010}.
Furthermore, the expected number of interactions per initial particle
(the \emph{multiplicity}) is much greater than one in many interaction
scenarios. At present, cutting-edge results are those in which the
final state contains only two additional particles, e.g.
double Compton scattering~%
\citep{seipt.prd.2012,mackenroth.prl.2013,king.pra.2015,dinu.prd.2019}
and trident pair creation~%
\citep{hu.prl.2010,ilderton.prl.2011,king.prd.2013,dinu.prd.2018,mackenroth.prd.2018},
due to the complexity of the calculations.

	\begin{figure}
	\centering
	\includegraphics[width=\linewidth]{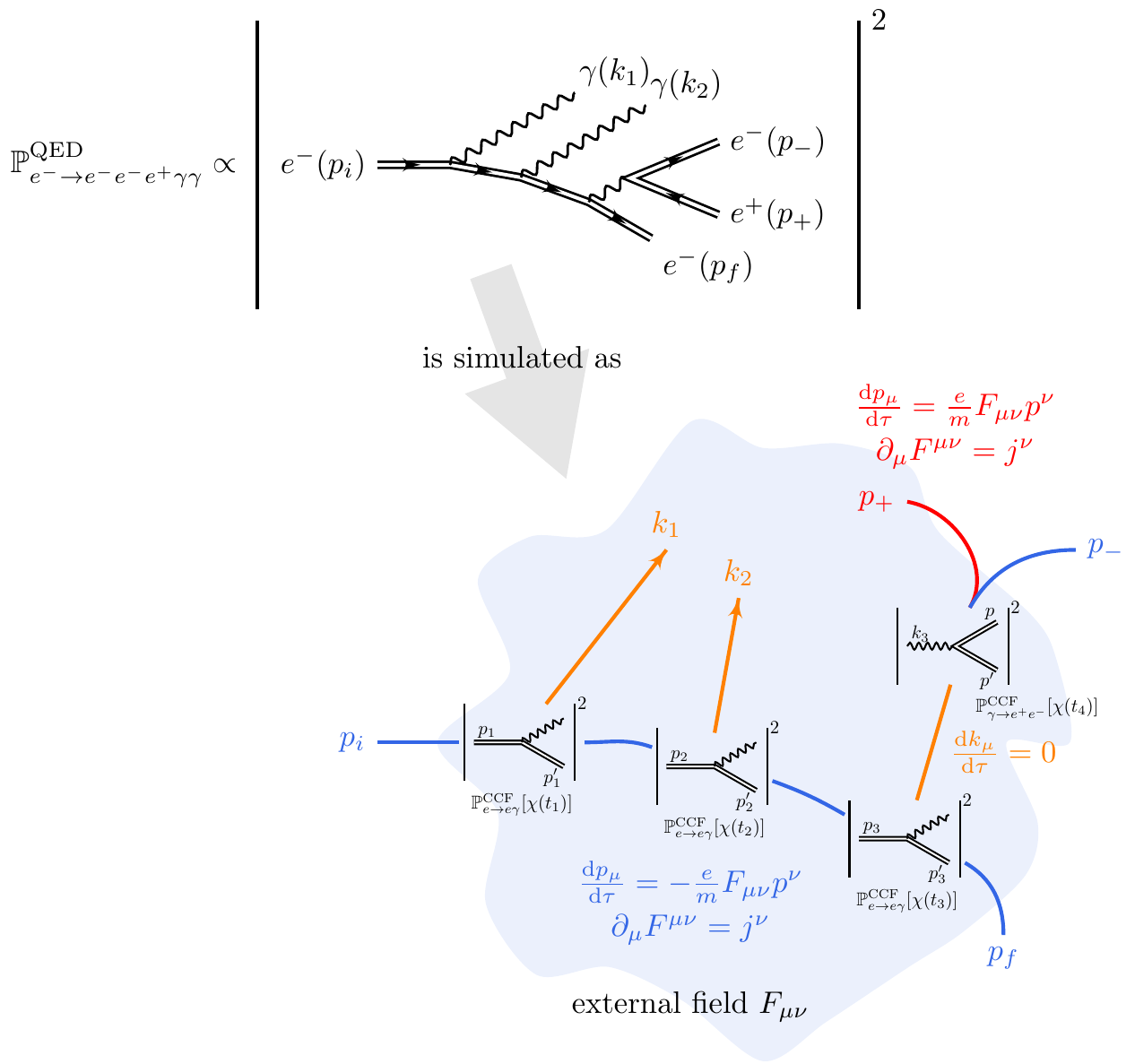}
	\caption{%
		A general strong-field QED interaction, featuring the emission
		and creation of multiple photons and electron-positron
		pairs, is simulated `semiclassically' by breaking it
		down into a chain of first-order processes
		(electrons, photons and positrons in blue, orange and red,
		respectively).
		Between these pointlike, instantaneous events, the
		particles follow classical trajectories guided by the
		Lorentz force: $\dot{p}_\mu = \pm e F_{\mu\nu} p^\nu / m$
		and $\dot{X}^\mu = p^\mu/m$,
		dots denoting differentiation with respect to proper time $\tau$.
		Modification of the external field $F_{\mu\nu}$
		is driven by the classical currents
		$j^\mu(x) = \pm (e/m) \int p^\mu(\tau) \delta^4[x - X(\tau)] \,\rmd\tau$.
		The probability rates and spectra for the first-order processes
		are those for a constant, crossed field, and depend on the
		local value of the quantum parameter $\chi(t) = \abs{F_{\mu\nu}[X(t)] p^\nu(t)} / (m \Ecrit)$.}
	\label{fig:Cascade}
	\end{figure}	

The need to overcome these issues has motivated the development of
numerical schemes that can model quantum processes at high multiplicity
in general electromagnetic fields. In this article we characterize these
schemes as `semiclassical', by virtue of the fact that they factorize
a QED process into a chain of first-order processes
that occur in vanishingly small regions linked by \emph{classically determined}
trajectories, as illustrated in \cref{fig:Cascade}.
The rates and spectra for the individual interactions
are calculated for the equivalent interaction in a constant, crossed
field, which may be generalized to an arbitrary field
configuration under certain conditions.
The first key result is that, at $a_0 \gg 1$, the formation length of
a photon (or an electron-positron pair) is much smaller than the length
scale over which the background field varies (see \cref{sec:Parameters})
and so emission may be treated as occurring instantaneously~\citep{ritus.jslr.1985}.
The second is that if $\chi^2 \gg \abs{\mathfrak{F}},\abs{\mathfrak{g}}$
and $\mathfrak{F}^2,\mathfrak{g}^2 \ll 1$, where
$\mathfrak{F} = (\vec{E}^2 - \vec{B}^2)/\Ecrit^2$ and
$\mathfrak{g} = \vec{E}\cdot\vec{B}/\Ecrit^2$ are the
two field invariants, the probability of a QED process is well
approximated by its value in a constant, crossed field:
$P(\chi,\mathfrak{F},\mathfrak{g}) \simeq P(\chi,0,0) + O(\mathfrak{F}) + O(\mathfrak{g})$
[see Appendix B of \citet{Baier}].
The combination of the two is called the \emph{locally constant, crossed
field approximation} (LCFA).
The first requires the laser intensity to be large, whereas the second
requires the particle to be ultrarelativistic and the background
to be weak (as compared to the critical field of QED).
We will discuss the validity of these approximations, and efforts to
benchmark them, in \cref{sec:Benchmarking}.

Within this framework, the laser-beam (or laser-plasma) interaction
is essentially treated classically, and quantum interactions such
as high-energy photon emission added by hand. The evolution of the
electron distribution function $\mathcal{F} = \mathcal{F}(t,\vec{r},\vec{p})$,
including the classical effect of the background field and stochastic photon emission,
is given by~\citep{elkina.prstab.2011,ridgers.jpp.2017}
	\begin{multline}
	\frac{\partial \mathcal{F}}{\partial t}
	+ \frac{\vec{p}}{\gamma m} \cdot \frac{\partial \mathcal{F}}{\partial \vec{r}}
	- e \left( \vec{E} + \frac{\vec{p}\times\vec{B}}{\gamma m} \right) \cdot \frac{\partial \mathcal{F}}{\partial \vec{p}}
	\\
	= -\mathcal{F} \int W_\gamma(\vec{p},\vec{k}') \,\rmd^3\vec{k}'
		+ \int \mathcal{F}(\vec{p}') W_\gamma (\vec{p}', \vec{p}' - \vec{p}) \,\rmd^3\vec{p}',
	\label{eq:EmissionVlasov}
	\end{multline}
where $W_\gamma(\vec{p},\vec{k}')$ is the probability rate for an electron
with momentum $\vec{p}$ to emit a photon with momentum $\vec{k}'$.
A direct approach to kinetic equations of this kind is to solve them
numerically~\citep{sokolov.prl.2010,bulanov.pra.2013,neitz.prl.2013},
or reduce them by means of a Fokker-Planck expansion in the limit
$\chi \ll 1$~\citep{niel.pre.2018}.
However, the most popular is a Monte Carlo implementation of the
emission operator [the right hand side of \cref{eq:EmissionVlasov}],
which naturally extends single-particle or particle-in-cell codes that
solve for the classical evolution of the distribution function
in the presence of externally prescribed, or self-consistent,
electromagnetic fields~\citep{duclous.ppcf.2011,elkina.prstab.2011}.
This method is discussed in detail in \citet{ridgers.jcp.2014,gonoskov.pre.2015},
so we only summarize it here for photon emission.

The electron distribution function is represented by an ensemble of
macroparticles, which represent a large number $w$ of real
particles ($w$ is often called the \emph{weight}). The trajectory of
a macroelectron between discrete emission events is determined
solely by the Lorentz force.
Each is assigned an
optical depth against emission $T = -\log(1-R)$ for pseudorandom
$0 \leq R < 1$, which evolves as $\frac{\rmd T}{\rmd t} = -W_\gamma$,
where $W_\gamma$ is the probability rate of emission,
until the point where it falls below zero.
Emission is deemed to occur instantaneously at this point and $T$ is reset.
The energy of the photon $\omega' = \abs{\vec{k}'}$ is pseudorandomly
sampled from the quantum emission spectrum
$\frac{\rmd N_\gamma}{\rmd \omega} = \omega^{-1} \frac{\rmd \Power_\text{q}}{\rmd \omega}(\chi,\gamma)$
[see \cref{eq:QuantumSpectrum}]
and the electron recoil determined by the conservation of momentum
$\vec{p} = \vec{p}' + \vec{k}'$ and the assumption that
$\vec{k}' \parallel \vec{p}$ if $\gamma \gg 1$.
If desired, a macrophoton with the same weight as the emitting macroelectron
can be added to the simulation. Electron-positron pair creation by
photons in strong electromagnetic fields is modelled in an analogous way
to photon emission~\citep{ridgers.jcp.2014,gonoskov.pre.2015}.

Thus there are two distinct descriptions of the electromagnetic field.
One component is treated as a classical field (in a PIC code, this would be discretized
on the simulation grid) and the other as a set of particles. In
principle this leads to double-counting; however, as we discussed
in \cref{sec:ModellingClassical}, the former lies at much lower frequency
than the photons that make up synchrotron emission, and has a
distinct origin in the form of externally generated fields (such as
a laser pulse) or the collective motion of a plasma.
Coherent effects are much less important for the
high-frequency components, which justifies describing them as
particles~\citep{gonoskov.pre.2015}.

\subsection{Benchmarking, extensions and open questions}
\label{sec:Benchmarking}

The validity of the simulation approach discussed in \cref{sec:ModellingQuantum}
relies on the assumption that a high-order QED process in a strong
electromagnetic background field may be factorized into a chain of first-order
processes, each of which is well approximated by the equivalent
process in a constant, crossed field. It is generally expected that
this reduction works in scenarios where $a_0 \gg 1$ and
$\chi^2 \gg \abs{\mathfrak{F}},\abs{\mathfrak{g}}$~\citep{ritus.jslr.1985,Baier}.
However, these asymptotic conditions do not give quantitative bounds
on the error made by semiclassical simulations. 
As these are the primary tool by which we predict radiation reaction effects
in high-intensity lasers, it is important that they are benchmarked
and that the approximations are examined.

	\begin{figure}
	\centering
	\includegraphics[width=\linewidth]{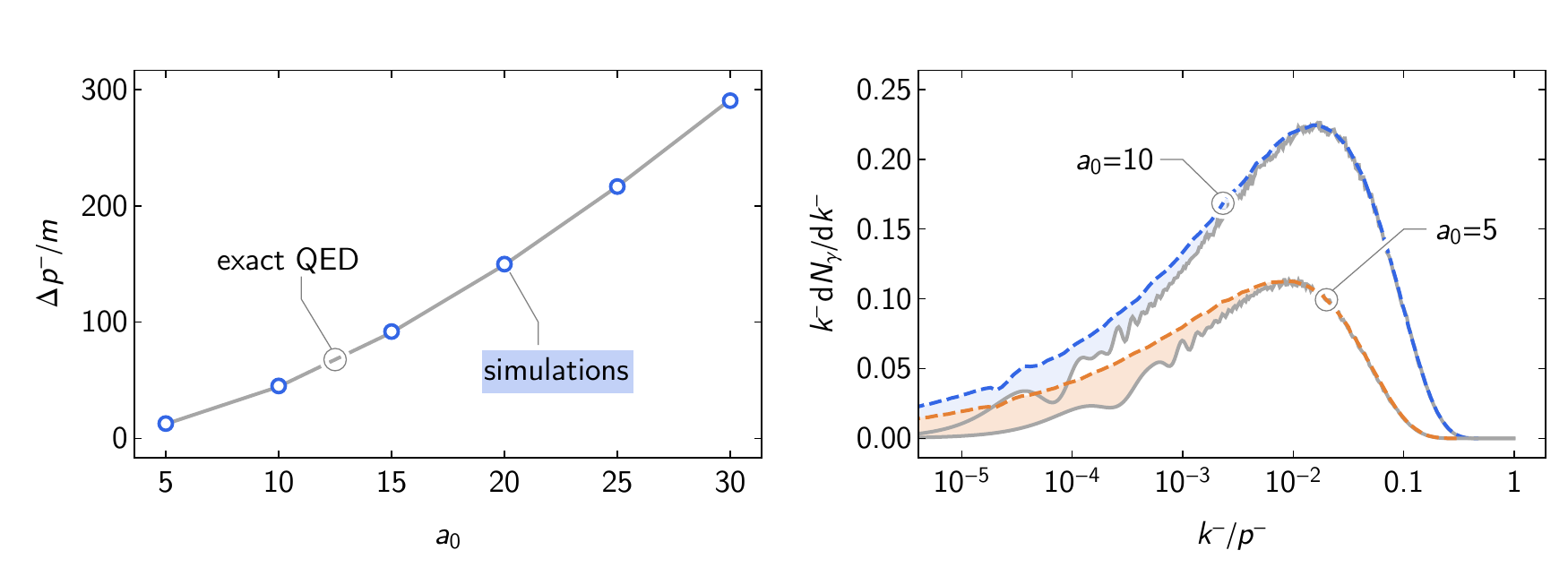}
	\caption{
		Comparison between exact QED (grey) and simulation results
		(blue and orange)
		for single nonlinear Compton scattering of an electron
		in a two-cycle, circularly polarized laser pulse:
		(left) the lightfront momentum loss as a function of
		laser amplitude $a_0$;
		and (right) exemplary photon spectra.
		Adapted from \citet{blackburn.pop.2018}.
		}
	\label{fig:Benchmarking}
	\end{figure}

One approach is to compare, directly, the predictions of strong-field
QED and simulations. We focus here on results for single nonlinear Compton
scattering~\citep{harvey.pra.2015,dipiazza.pra.2018,blackburn.pop.2018},
the emission of one and only one photon in the interaction of
an electron with an intense, pulsed plane EM wave, by virtue of its
close relation to radiation reaction.
It is shown that the condition $a_0 \gg 1$ is necessary, but not sufficient,
for the applicability of the LCFA: we also require that $a_0^3 / \chi \gg 1$
for interference effects to be suppressed~\citep{dinu.prl.2016}.
These interference effects are manifest in the low-energy part of the
photon emission spectrum $x = \omega' / (\gamma m) < \chi/a_0^3$,
as the formation length for such photons is comparable in size to
the wavelength of the background field.
Semiclassical simulations strongly overestimate the number of photons
emitted in this part of the spectrum because they exclude nonlocal
effects~\citep{harvey.pra.2015,dipiazza.pra.2018}.
Nevertheless, they are much more accurate with respect to the
total energy loss (and therefore to radiation reaction), because
this depends on the \emph{power} spectrum, to which the low-energy
photons do not contribute significantly~\citep{blackburn.pop.2018}.
This is shown in \cref{fig:Benchmarking}, which compares the
predictions of exact QED and semiclassical simulations for
an electron with $p^-_0 / m \simeq 2 \gamma_0 = 2000$ colliding
with a two-cycle laser pulse with normalized amplitude $a_0$ and
wavelength $\lambda = 0.8~\micron$.
There is remarkably good agreement between the two even for
$a_0 = 5$.

An additional point of comparison in \citet{blackburn.pop.2018}
is the number of photons absorbed from the background field
in the process of emitting a high-energy photon. This transfer
of energy from the background field to the electron is required
by momentum conservation. Without emission, there would be no
such transfer of energy. This is consistent with the classical
picture, in which plane waves do no work in the absence of
radiation reaction.
Strong-field QED calculations depend crucially on the fixed
nature of the background field; however, for single nonlinear Compton
scattering, near-total depletion of the field is predicted at
$a_0 \gtrsim 1000$~\citep{seipt.prl.2017}. The theory
must therefore allow for changes to the background~\citep{ilderton.prd.2018}.
Within the semiclassical approach, depletion is accounted for
by the action of the classical currents
through the $\vec{j}\cdot\vec{E}$ term in Poynting's theorem.
Quantum effects are manifest in how photon emission (and pair creation),
modify those classical currents, as illustrated in \cref{fig:Cascade}.
In \citet{blackburn.pop.2018}, the classical work done on the
electron is shown to agree well with the number of absorbed
photons predicted by exact QED. This is consistent with the
results of \citet{meuren.prd.2016}, which indicate that the
`classical' dominates the `quantum' component of depletion,
the latter associated with absorption over the formation length,
if $a_0 \gg 1$.

The failure of the semiclassical approach to reproduce the
low-energy part of the photon spectrum arises from the localization
of emission. Most notably, the number spectrum
$\frac{\rmd N_\gamma}{\rmd \omega} =
\omega^{-1} \frac{\rmd \Power_\text{q}}{\rmd \omega}(\chi,\gamma)$
[see \cref{eq:QuantumSpectrum}]
diverges as $\omega^{-2/3}$ as $\omega \to 0$.
This can be partially ameliorated by the use of
emission rates that take nonlocal effects into account.
\Citet{dipiazza.pra.2019} suggest replacing the LCFA spectrum
in the region $x \lesssim \chi/a_0^3$ with the equivalent, finite, result
for a monochromatic plane wave, which they adapt for use
in arbitrary electromagnetic field configurations.
\Citet{ilderton.pra.2019} propose an approach based on formal
corrections to the LCFA, in which the emission rates depend
on the field \emph{gradients} as well as magnitudes.

While the studies discussed above have given insight into the
limitations of the LCFA, they do not examine the applicability
of the factorization shown in \cref{fig:Cascade}, as this
requires by definition the calculation of a higher order
QED process. At the time of writing, there are no direct
comparisons of semiclassical simulations and strong-field QED
for either double Compton scattering (emission of two photons)
or trident pair creation (emission of a photon which decays
into an electron-positron pair). Factorization, also called
the \emph{cascade approximation}, has been examined directly
within strong-field QED for the trident process in a constant
crossed field~\citep{king.prd.2013} and in a pulsed plane
wave~\citep{dinu.prd.2018,mackenroth.prd.2018}.
In the latter it is shown that at $a_0 = 50$ and an electron
energy of 5~GeV, the error is approximately one part in a thousand.

The dominance of the cascade contribution makes it important
to consider whether the propagation of the electron between
individual tree-level process, as shown in \cref{fig:Cascade},
is done accurately. In the standard implementation, this is
done by solving a classical equation of motion including
only the Lorentz force~\citep{ridgers.jcp.2014,gonoskov.pre.2015}.
The evolution of the electron's spin is usually neglected
and emission calculated using unpolarized rates, such as \cref{eq:QuantumSpectrum}.
\Citet{king.pra.2015} show that the accuracy of modelling
double Compton scattering in a constant crossed field
as two sequential emissions with unpolarized rates is
better than a few per cent.
There are, however, scenarios, where the spin degree of
freedom influences the dynamics to a larger degree.
Modelling these interactions with semiclassical simulations
requires spin-resolved emission rates~\citep{SokolovTernov,seipt.pra.2018}
and an equation of motion for the electron spin~\citep{thomas.nature.1926,bargmann.prl.1959}.
In a rotating electric field, as found at the magnetic node
of an electromagnetic standing wave~\citep{bell.prl.2008},
where the spin does not precess between emissions,
the asymmetric probability of emission between different
spin states leads to rapid, near-complete polarization of the electron
population~\citep{delsorbo.pra.2017,delsorbo.ppcf.2018}.
Similarly, an electron beam interacting with a linearly polarized
laser pulse can acquire a polarization of a few per cent~\citep{seipt.pra.2018}.
To make this larger, it is necessary to break the symmetry in the
field oscillations, which can be accomplished by introducing a small
ellipticity to the pulse~\citep{li.prl.2019}, or by superposition
of a second colour~\citep{song.pra.2019,chen.prl.2019,seipt.pra.2019}.

A more fundamental limitation on the applicability of the LCFA
is that the emission rates are calculated at tree level only.
The importance of loop corrections to the strong-field QED vertex
grows as $\alpha \chi^{2/3}$ in a constant, crossed
field~\citep{morozov.jetp.1981}, leading to speculation that
$\alpha \chi^{2/3}$ is the `true' expansion parameter of
strong-field QED~\citep{narozhny.prd.1980}. When $\chi \simeq 1600$,
this parameter becomes of order unity and the meaning of a
perturbative expansion in the dynamical electromagnetic field
breaks down. The recent review by \citet{fedotov.jpc.2017} has
prompted renewed interest in this regime; recent calculations of
the one-loop polarization and mass operators~\citep{podszus.prd.2019}
and photon emission and helicity flip~\citep{ilderton.prd.2019}
in a general plane-wave background have confirmed that the
power-law scaling of radiative corrections pertains strictly
to the high-intensity limit $a_0^3 / \chi \gg 1$. In the high-energy
limit, radiative corrections grow logarithmically, as in
ordinary (i.e., non-strong-field) QED~\citep{podszus.prd.2019,ilderton.prd.2019}.

The difficulty in probing the regime $\alpha \chi^{2/3} \gtrsim 1$
is the associated strength of radiative energy losses,
which suppress $\gamma$ and so $\chi$~\citep{fedotov.jpc.2017}.
Overcoming this barrier at the desired $\chi$ requires the
interaction duration to be very short.
The beam-beam geometry proposed by \citet{yakimenko.prl.2019}
exploits the Lorentz contraction of the Coulomb field of a
compressed (100~nm), ultrarelativistic (100~GeV) electron beam,
which is probed by another beam of the same energy.
In the laser-electron-beam scenario considered by \citet{blackburn.njp.2019},
collisions at oblique incidence are proposed for reaching
$\chi \gtrsim 100$, exploiting the fact that the diameter of
a laser focal spot is typically much smaller than the
duration of its temporal profile.
Even higher $\chi$ is reached in the combined laser-plasma,
laser-beam interaction proposed by \citet{baumann.ppcf.2019}.
While it seems possible to approach the fully nonperturbative regime
experimentally, albeit for extreme collision parameters, there is
no suitable theory at $\alpha \chi^{2/3} \gtrsim 1$, and quantitative
predictions are lacking in this area.

\section{Experimental geometries, results and prospects}

\subsection{Geometries}
\label{sec:Geometries}

	\begin{figure}
	\centering
	\includegraphics[width=0.6\linewidth]{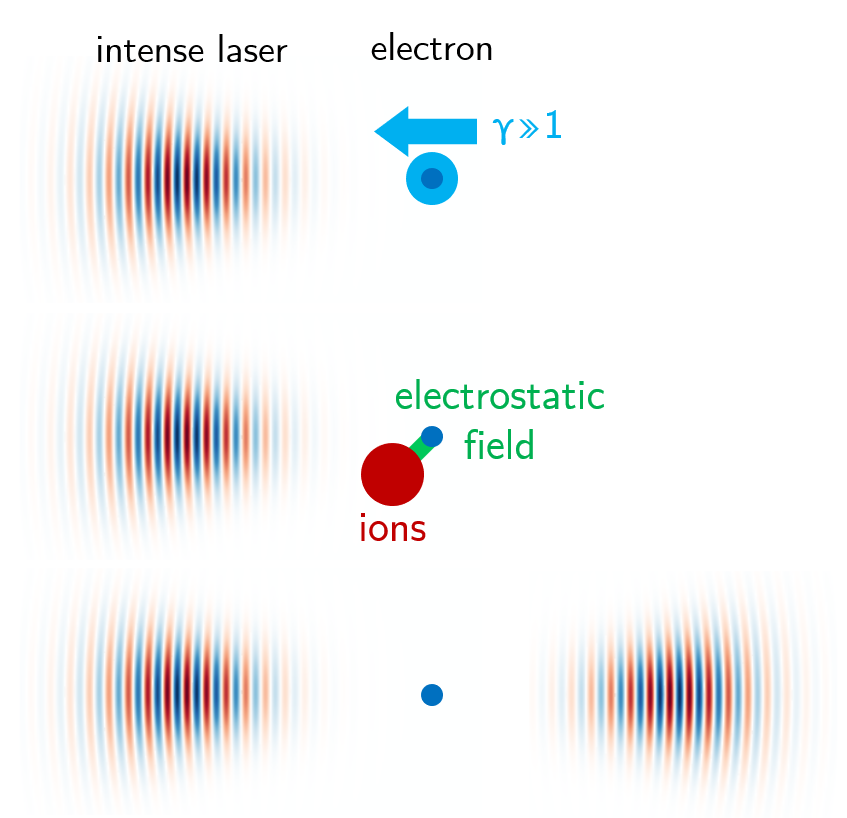}
	\caption{Tightly focussed laser pulses can ponderomotively
				expel electrons from the region of highest intensity,
				suppressing the onset of radiation-reaction and
				quantum effects.
				There are three typical experimental geometries
				that ensure that energetic electrons are embedded
				in strong EM fields as desired:
				(top) laser--particle-beam,
				(centre) laser-plasma,
				and (bottom) laser-laser.
				}
	\label{fig:Geometries}
	\end{figure}
	
It may be appreciated that the radiation-reaction and quantum
effects under consideration here, as particle-driven processes,
can only become important if electrons or positrons are
actually embedded within electromagnetic fields of suitable strength.
However, the estimates in \cref{sec:Parameters} were made for a plane
EM wave, in which case the electron is guaranteed to interact
with the entire wave, including the point of highest intensity.
In reality, such intensities are reached by compressing
the laser energy into ultrashort pulses~\citep{strickland.oc.1985}
that are focussed to spot sizes close to the
diffraction limit~\citep{bahk.ol.2004,sung.ol.2017,kiriyama.ol.2018}.
The steep spatiotemporal gradients in intensity that result
mean that laser pulses can ponderomotively expel electrons from
the focal region, in both vacuum~\citep{malka.prl.1997,thevenet.np.2016}
and plasma~\citep{esarey.rmp.2009}, curtailing the interaction
long before the particles experience high $a_0$ or $\chi$.

The literature contains many possible experimental configurations
designed to explore or exploit radiation reaction and quantum
effects. These configurations can be divided, broadly, into
three categories, based on how they ensure the spatial
coincidence between particles and strong fields.
\Cref{fig:Geometries} illustrates the three categories.
In the first (\emph{laser--particle-beam}),
the electrons are accelerated to ultrarelativistic
energies before they encounter the laser pulse. The effective `mass
increase' makes the beam rigid and so it passes through the entirety of
the laser pulse, avoiding substantial deflection and ensuring
that it is exposed to the strongest electromagnetic fields.
Concretely, the ponderomotive force is suppressed at high $\gamma$:
$\rmd \avg{\vec{p}}/\rmd t = -m \vec{\nabla} \avg{a^2} / (2\avg{\gamma})$,
where $\avg{\cdot}$ denotes a cycle-averaged quantity~\citep{quesnel.pre.1998}.
It should be noted that it is possible for radiation reaction
to amplify this force to the point that it can prevent an
arbitrarily energetic electron from penetrating the laser
field~\citep{zhidkov.prstab.2014,fedotov.pra.2014}; however,
this requires $a_0 \gtrsim 300$, far in excess of what it is
available at present.
In today's high-intensity lasers,
ultrarelativistic electrons can reach the nonlinear quantum
regime $\chi \sim 1$ even for $a_0 \sim 10$ (see \cref{fig:a0-chi}).

In the second (\emph{laser-plasma}), the electrons are
electrostatically bound to a population of ions, which
are substantially more massive and therefore less mobile.
Large-scale displacement of the electrons away from the laser
fields is then suppressed by the emergence of plasma fields.
If the plasma is overdense, i.e. opaque to the laser light,
then only electrons in a thin layer near the surface
experience the full laser intensity and are accelerated
to relativistic energies.
However, the high density of electrons in this region means
that a significant fraction of the laser energy is converted to
high-energy radiation, leading to, for example,
dense bursts of $\gamma$ rays and positrons~\citep{ridgers.prl.2012,brady.prl.2012},
reduced efficiency of ion acceleration~\citep{tamburini.njp.2010}
and the generation of long-lived quasistatic magnetic fields~\citep{liseykina.njp.2016}.
If the target is close to underdense, by contrast,
the laser can propagate through the plasma bulk and the interaction
is volumetric in nature.
The combination of laser and induced plasma fields, as well as
radiation reaction, leads to confinement and acceleration of the
electrons, and copious emission of radiation~\citep{stark.prl.2016,zhu.nc.2016,vranic.ppcf.2018}.

Finally, electrons can be trapped in the collision of more
than one laser pulse (\emph{laser-laser}), where they interact
with an electromagnetic standing, rather than travelling,
wave~\citep{bulanov.prl.2010}.
Radiation reaction induces a rich set of dynamics in this
configuration~\citep{li.prl.2014,gonoskov.prl.2014,%
esirkepov.pla.2015,kirk.ppcf.2016}.
The fact that standing waves can do work in reaccelerating
the particles after they recoil means that, at intensities
$\gtrsim 10^{24}~\Wcmsqd$, the emitted photons seed avalanches
of electron-positron pair creation~\citep{bell.prl.2008};
this intensity threshold is lowered in suitable multibeam
setups~\citep{gelfer.pra.2015,vranic.ppcf.2017,gong.pre.2017}.
The case of optimal focussing is achieved in a dipole
field~\citep{bassett.oa.1986}, where the peak $a_0 \simeq
780 \Power^{1/2} [\text{PW}]$~\citep{gonoskov.pra.2012}.
Such extreme intensities, at moderate power, are the reason this
configuration has been studied as means of high-energy
photon production~\citep{gonoskov.prx.2017,magnusson.prl.2019}.

It is important to note that the distinction between the
three categories defined here is not absolute. Mixing between
them occurs in, for example, the interaction of a linearly
polarized laser pulse with relativistically underdense plasma: here
\emph{re-injected electron synchrotron emission}, the radiation
emission when electrons are pulled backwards into the oncoming
laser by a charge-separation field~\citep{brady.prl.2012},
exhibits features of both the `laser-plasma' and `laser-beam'
geometries.
Furthermore, the exponential growth of particle number in a
QED cascade driven by multiple laser pulses can create an
electron-positron plasma of sufficient density to shield
the interior from the laser pulse~\citep{grismayer.pop.2016},
leading to a transition between the `laser-laser' and `laser-plasma'
categories.
Twin-sided illumination of a foil has features of both
\emph{ab initio}~\citep{zhang.njp.2015}.

\subsection{`All-optical' colliding beams}
\label{sec:AllOptical}

This paper focuses on the first of the three
configurations discussed in \cref{sec:Geometries},
\emph{laser--particle-beam},
for the reason that it allows $\chi > 0.1$ to be reached
at lower $a_0$ than would be required in a laser-plasma
or laser-laser interaction.
As is shown in \cref{fig:a0-chi} and by \cref{eq:Chi},
given a 500-MeV electron beam, quantum effects on radiation
reaction can be reached even at an intensity of $10^{21}~\Wcmsqd$.
The colliding beams geometry therefore represents
a promising first step towards experimental exploration of the
radiation-dominated or nonlinear quantum regimes.

Thus far we have not specified the source of ultrarelativistic
electrons. The theoretical description of the interaction does
not depend on the source, of course, but it is of immense practical
importance. Furthermore, the characteristics of the source (its
energy, bandwidth, emittance, etc.) are key determining factors
in the viability of measuring radiation-reaction or quantum effects.
For example, the fact that electron beam energy spectra are
expected to broaden due to stochastic effects makes the variance
of the spectrum, $\sigma^2$, an attractive signature of the
quantum nature of radiation reaction~\citep{neitz.prl.2013,vranic.njp.2016}.
However, such broadening can occur classically in the interaction
of an electron beam with a focussed laser pulse, because
components of the beam can encounter different intensities and therefore
lose different amounts of energy~\citep{harvey.prab.2016}.
Thus a crucial role is played by the initial energy spread and
size of the incident electron beam~\citep{samarin.jmo.2017}.

In fact, it was pioneering experiments with a conventional,
radio-frequency (RF), linear accelerator that provided the first
demonstration of nonlinear quantum effects in a strong
laser field:
nonlinearities were measured in Compton scattering~\citep{bula.prl.1996}
and Breit-Wheeler electron-positron pair creation~\citep{burke.prl.1997}
in Experiment 144 at the Stanford Linear Accelerator (SLAC) facility.
In this experiment, the $46.6$~GeV electron beam was collided with
a laser pulse of intensity $1.3\times10^{18}~\Wcmsqd$, duration $1.4$~ps
and wavelength $527$~nm ($a_0 \simeq 0.4$, $\chi = 0.3$)~\citep{bamber.prd.1999}.
The pair creation mechanism was reported to be the multiphoton
Breit-Wheeler process, as $n = 4$ laser photons were required in
addition to each multi-GeV $\gamma$ ray (emitted by the electron
in Compton scattering) to overcome the mass threshold.\footnote{%
More recent analysis,
in which the two stages of photon emission and pair creation
are treated within a unified framework using strong-field QED,
thereby including the direct, `one-step', contribution,
indicates that the experiment did, in fact, observe the onset
of nonperturbative effects~\citep{hu.prl.2010}, see also~\citep{ilderton.prl.2011}.}
In total, $106 \pm 14$ positrons were observed
over the series of 22,000 laser shots. The yield was strongly limited
because, even though the electron energy was sufficient to reach a
quantum parameter $\chi \sim 0.3$, in the regime $a_0 \ll 1$,
the pair creation probability is suppressed as $a_0^{2n}$, where
$n$, the number of participating photons, was found to be $n \simeq 5$~\citep{burke.prl.1997}.
Similarly, the photon emission process was weakly nonlinear, with
harmonics of the fundamental Compton energy up to $n = 4$
observed~\citep{bula.prl.1996}.
At the time of writing, this experiment had yet to be repeated at
a conventional linear accelerator, though concrete proposals
have now been made to do so at DESY~\citep{abramowicz.arxiv.2019}
and FACET-II~\citep{meuren.exhilp.2019}. While the electron beams
will be less energetic ($17.5$ and $10$~GeV respectively), the
laser intensity will be higher ($2 \lesssim a_0 \lesssim 10$),
so the transition from the multiphoton to the tunnelling regimes
could be explored.

One of the challenges that must be overcome in realizing these
experiments is that, as discussed in \cref{sec:Geometries},
lasers reach high intensity by focussing and compressing energy into
a small spatiotemporal volume.
Thus the region in which the electromagnetic fields are strong
is only a few microns in radius, assuming diffraction-limited
focussing and optical drivers ($\omega \sim 1$~eV), which
is much smaller than the size of the focussed electron beam
from a conventional accelerator.
This limits the number of electrons that interact with the
laser, reducing the relevant signal, as well as making the
alignment and timing of the beams more difficult~\citep{samarin.jmo.2017}.
In the `all-optical' geometry, these are overcome by using
a dual-laser setup~\citep{bulanov.nima.2011}: one laser provides
the high-intensity `target', and the other is used to accelerate
electrons in a plasma wakefield.

Laser-driven wakefield acceleration has undergone remarkable
progress over the last two decades: from the first
quasi-monoenergetic relativistic beams~\citep{mangles.nature.2004,%
geddes.nature.2004,faure.nature.2004}, they now produce
electron beams with near- to multi-GeV energies~\citep{kneip.prl.2009,%
wang.nc.2013,gonsalves.prl.2019}.
Briefly, an intense laser pulse travels through a low-density plasma,
exciting, via its ponderomotive effect, a trailing nonlinear
plasma wave that traps and accelerates electrons~\citep{esarey.rmp.2009}.
As the medium is a plasma, already ionized and therefore immune
to electrostatic breakdown, the accelerating gradients are much
higher than in a conventional RF accelerator: GeV energies
can be reached in only a few centimetres of propagation.
Furthermore, as the size of the accelerating structure is only
a few microns (at typical plasma densities $n_e \sim 10^{18}~\text{cm}^{-3}$,
the plasma wavelength $\sim 20~\micron$), the electron beams
produced in wakefield acceleration are similarly micron-scale,
with durations of order 10~fs.

Besides the high energy and the small size of the electron beam,
we have the intrinsic synchronizations of the electron beam with
the accelerating laser pulse, and of the accelerating laser
pulse with the target laser pulse, if the two emerge from amplifier
chains that are seeded by the same oscillator.
Thus the `all-optical' laser-electron-beam collision is promising
as a compact source of bright, ultrashort bursts of high-energy
$\gamma$ rays~\citep{chen.prl.2013,sarri.prl.2014,yan.np.2017}.
Now, with advances in laser technology, a multibeam facility is capable
of reaching the radiation-reaction and nonlinear quantum regimes.
Recently two such experiments were performed using the Gemini
laser at the Rutherford Appleton Laboratory~\citep{cole.prx.2018,poder.prx.2018},
a dual-beam system that delivers twin synchronized pulses of
duration $45$~fs and energy $\sim 10$~J, with a peak $a_0 \simeq 20$:
we discuss these experiments in detail in \cref{sec:RecentResults}.
Upcoming laser facilities, such as Apollon~\citep{papadopoulos.hpl.2016}
or ELI~\citep{weber.mre.2017,gales.rpp.2018}, aim for laser-electron
collisions at even higher intensity: see, for example, \citet{lobet.prab.2017}
for simulations of dual-beam interactions at $a_0 \simeq 200$.

Not all high-intensity laser facilities have dual-beam capability.
An alternative all-optical configuration, introduced by \citet{taphuoc.np.2012},
employs a \emph{single} laser pulse as accelerator and target:
a foil is placed at the end of a gas jet, into which a laser is focussed
to drive a wakefield and accelerate electrons; when the laser pulse
reaches the foil, it is reflected from the ionized surface back onto the
trailing electrons.
This guarantees temporal and spatial overlap of the two beams, but
precludes the possibility of separately optimizing the two laser pulses;
in \citet{taphuoc.np.2012} the electron energies $\simeq 100$~MeV
and the peak $a_0 \simeq 1.5$, so radiation reaction effects were negligible.
Simulations of similar single-pulse geometries predict the efficient
production of multi-MeV photons at $a_0 > 50$~\citep{huang.njp.2019} and
electron-positron pairs at $a_0 \gtrsim 300$~\citep{liu.ppcf.2016,gu.cp.2018}.

\subsection{Recent results}
\label{sec:RecentResults}

	\begin{figure}
	\centering
	\includegraphics[width=0.7\linewidth]{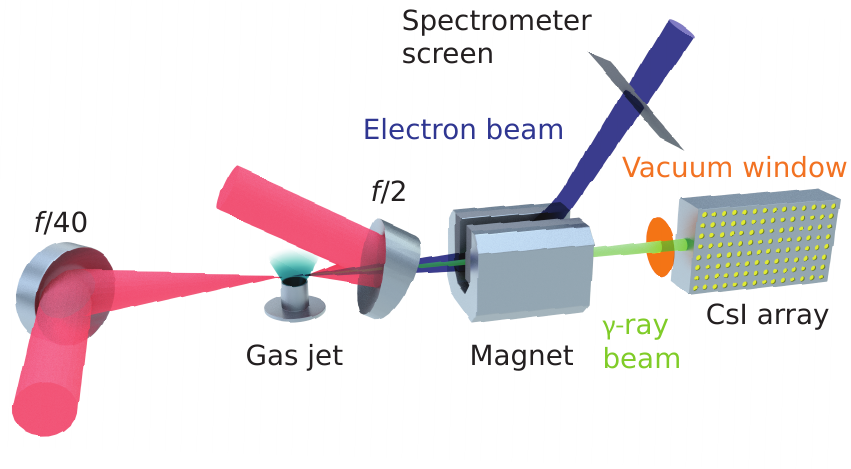}
	\caption{
		Layout of an all-optical colliding-beams experiment.
		A hole in the short-focal-length ($F/2$) optic allows
		for counterpropagation of the electron beam, which
		is accelerated by a laser wakefield in a gas jet,
		and the high-intensity laser pulse.
		The decelerated electrons, the $\gamma$ rays they
		emit in the collision, and the accelerating
		laser pulse, pass through this hole before being
		blocked or diagnosed as appropriate.
		The collision is timed to occur close to the rear
		of the gas jet (on the right-hand side, as viewed in the figure),
		before the electron beam can diverge significantly,
		which maximizes overlap between the beams.
		Reproduced from \citet{cole.prx.2018}.
	}
	\label{fig:CollidingBeams}
	\end{figure}

The Gemini laser of the Central Laser Facility (Rutherford Appleton Laboratory, UK)
is a petawatt-class dual-beam system~\citep{hooker.jpf.2006}, well-suited
for the all-optical colliding beams experiments discussed in \cref{sec:AllOptical}.
It delivers two, synchronized, linearly polarized laser pulses of duration $45~\fs$,
energy 10~J and wavelength $0.8~\micron$.
Available focussing optics include long-focal-length mirrors ($F/20$ and $F/40$)
for laser-wakefield acceleration and, most importantly, a short-focal-length ($F/2$)
off-axis parabolic mirror with an $F/7$ hole in its centre~\citep{cole.prx.2018,poder.prx.2018}.
The latter allows for direct counterpropagation of the two laser beams,
the geometry in which $\chi$ is largest (see \cref{eq:ChiGeneral}):
the more weakly focussed laser that drives the wakefield passes through
the hole and is subsequently blocked, avoiding backreflection in the amplifier chains;
the accelerated electron beam, and any radiation produced in the
collision with the tightly focussed laser, can pass through to reach the diagnostics.
Both the experiments that will be described in this section used this geometry,
which is illustrated in \cref{fig:CollidingBeams}, but with different electron
acceleration stages.

In \citet{cole.prx.2018}, the accelerating laser pulse was focussed onto the
leading edge of a supersonic helium gas jet, producing a ${\sim}15$~mm plasma
acceleration stage with peak density $n_e \simeq 3.7 \times 10^{18}~\text{cm}^{-3}$.
The use of a gas jet allowed the second laser pulse to be focussed close to
the point where the electron beam emerges from the plasma (at the rear edge),
so the collision between electron beam and laser pulse
took place when the former was much smaller than the latter (approximately
$1~\micron^2$ rather than $20~\micron^2$, which includes the effect of
a systematic time delay between the two).
The advantages of using a gas cell, as in \citet{poder.prx.2018}, are the
higher electron beam energies and significantly better shot-to-shot stability.
However, in this case, the second laser pulse must be focussed further
downstream of the acceleration stage (approximately 1~cm), by which point the electron
beam has expanded to become comparable in size to the laser.
Thus full 3D simulations were required for theoretical modelling of
the interaction, whereas 1D (plane-wave) simulations were sufficient
in \citet{cole.prx.2018}.

Fluctuations in the pointing and timing of the two lasers, as well as
systematic drifts in the latter, mean that the overlap between
electron beam and target laser pulse varies from shot to shot.
It is helpful, therefore, to gather as large a dataset as possible
(with the second, high-intensity, laser pulse both on and off),
in which case high-repetition-rate laser systems are at a clear
advantage. However, this is not nearly so important as being able
to identify `successful' collisions when they occur; even a small
set of collisions ($N \sim 10$) can provide statistically
significant evidence of radiation reaction when this is done.
This speaks to the importance of measuring both the electron and
$\gamma$-ray spectra on a shot-to-shot basis; identifying coincidences
between the two provides stronger evidence of radiation reaction
than could be obtained by either alone.

In \citet{cole.prx.2018}, successful shots were distinguished by
measuring the total signal in the $\gamma$-ray detector
$S_\gamma \propto N_e a^2 \avg{\gamma^2}$ (background-corrected),
where $N_e$ is the total number of electrons in the beam,
$\avg{\gamma^2}$ their mean squared Lorentz factor, and $a$ an
overlap-dependent, effective value for $a_0$
(the former two can be extracted from the measured electron spectra).
Over a sequence of 18 shots (eight \emph{beam-on}, i.e. with
the $f/2$ beam on, ten \emph{beam-off}), four were measured with a
normalized CsI signal, $\hat{S}_\gamma = S_\gamma / (N_e \avg{\gamma^2}) \propto a^2$,
four standard deviations above the background level.
These four also had electron beam energies below 500~MeV 
(as identified by a strong peak feature in the measured spectra),
whereas the ten beam-off shots had a mean energy of $550 \pm 20$~MeV.
The probability of measuring four or more beams with energy below
500~MeV in a sample of eight, given this fluctuation alone,
is approximately 10\%.
However, the probability that four beams have this lower energy
\emph{and} a significantly higher $\gamma$-ray signal is the
considerably smaller 0.3\%.

	\begin{figure}
	\centering
	\includegraphics[width=0.49\linewidth]{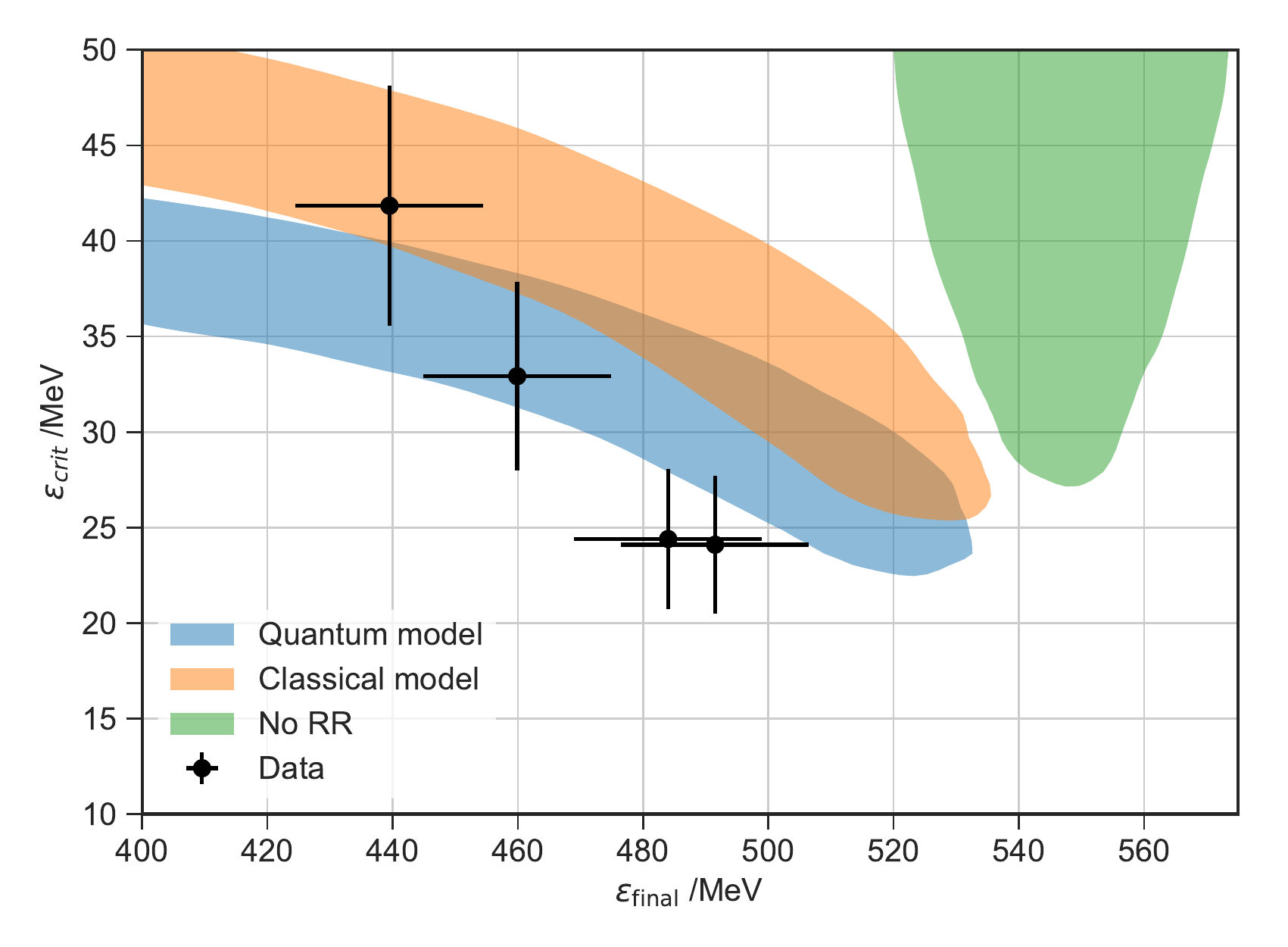}
	\includegraphics[width=0.49\linewidth]{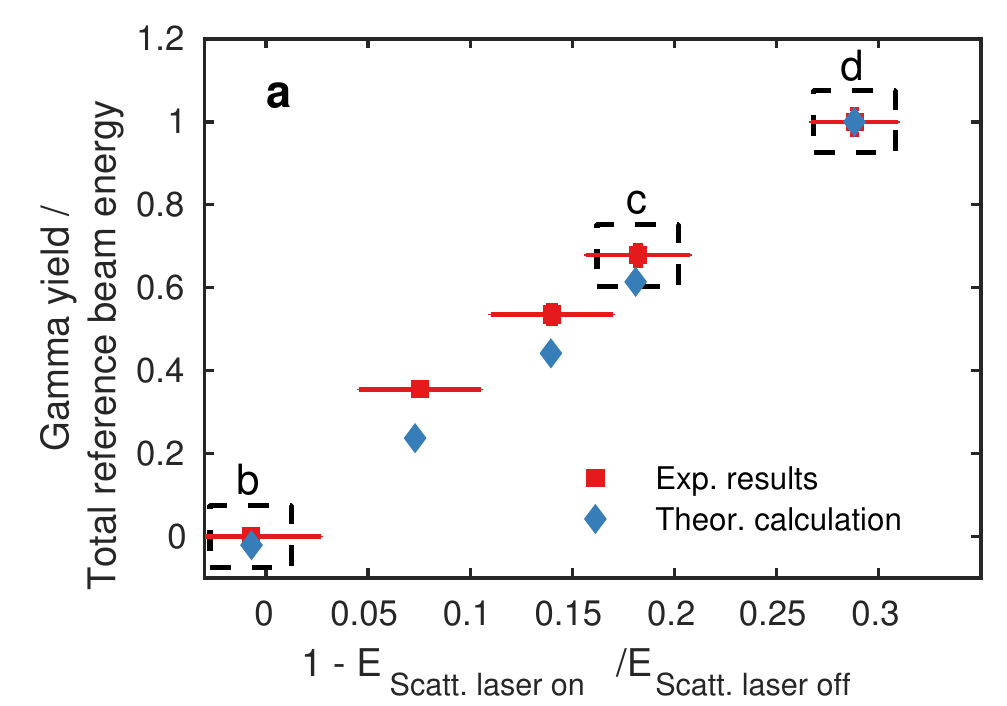}
	\caption{
		Experimental evidence of radiation reaction:
		(left) in \citet{cole.prx.2018},
		the post-collision electron beam energies and critical
		energies of the $\gamma$-ray spectra (black points),
		are consistent with theoretical simulations
		that include radiation reaction, with slightly better
		agreement for the stochastic model;
		(right) in \citet{poder.prx.2018},
		the fractional reduction in the total electron beam energy
		is correlated with the total $\gamma$-ray signal,
		with the best agreement with theory given by the
		`modified classical' model (see \cref{sec:ModellingClassical}).
		Details are given in the main text.
		}
	\label{fig:Cole}
	\end{figure}

Statistically significant evidence of radiation reaction
was obtained by correlating the electron beam energy
with the \emph{critical energy} of the $\gamma$-ray spectrum
$\varepsilon_\text{crit}$, a parameter characterizing the
hardness of the spectrum. This was accomplished by fitting the
depth-resolved scintillator output to a parametrized spectrum
$dN_\gamma/d\omega
\propto \omega^{-2/3} \exp(-\omega / \varepsilon_\text{crit})$,
having first characterized its response to monoenergetic photons
in the energy range $2 < \omega [\text{MeV}] < 500$
with \textsc{Geant4} simulations (see details in \citet{behm.rsi.2018}).
This choice of parametrization approximately reproduces
a synchrotron-like spectral shape, with an exponential
rollover at high energy and a scaling like $\omega^{-2/3}$ at
low energy.
The four successful shots demonstrate a negative correlation
between the final electron energy and $\varepsilon_\text{crit}$,
as is shown in \cref{fig:Cole}; this is consistent with
radiation-reaction effects, as the hardest photon spectra should
come from electron beams that have lost the most energy.
The probability to observe this negative correlation and
to have electron energy lower than 500~MeV on all four
successful shots is 0.03\%, which qualifies, under the usual
three-sigma threshold, as evidence of radiation reaction.

Simulations of the collision confirmed that the critical energies
and electron energy loss were consistent with theoretical expectations
of radiation reaction.
The coloured regions in \cref{fig:Cole} give the areas in which
68\% (i.e. one sigma) of results would be found for a large
ensemble of `numerical experiments', given the measured fluctuations
in the pre-collision electron energy spectra and the collision $a_0$,
and under specific models of radiation reaction.
The results exclude the `no RR' model, in which the electrons
radiate, but do not recoil. They are more
consistent with the stochastic, quantum model discussed in
\cref{sec:ModellingQuantum} than the deterministic, classical
model of Landau-Lifshitz: however, it is important to note
that both models are consistent with the data at the two-sigma level.
Subsequent analysis has confirmed that the `modified classical' model
discussed in \cref{sec:ModellingClassical}, which includes
the quantum suppression factor $g(\chi)$, given in \cref{eq:Gaunt},
but not the stochasticity of emission, gives practically
the same region as the quantum model~\citep{cole.pc.2018},
as is stated in \citet{cole.prx.2018}.
This is because the electron beam energy effectively parametrizes
the mean of the spectrum, the evolution of which depends only
on $g(\chi)$ according to \cref{eq:MuEvolution}; to see
stochastic effects, we must consider instead the width
of the distribution~\citep{neitz.prl.2013,vranic.njp.2016,ridgers.jpp.2017}.
Given electron beams with narrower initial energy spectra,
it would be possible to identify stochastic effects (or
their absence) by correlating the mean and variance of
the final electron energy spectra~\citep{arran.ppcf.2019}.

Evidence of radiation reaction was also obtained in the experiment
reported by \citet{poder.prx.2018}, by a complementary form of analysis.
The total CsI signal was used to discriminate successful collisions:
the signal normalized to the total energy in
a reference (beam-off) shot $S_\gamma / (N_e \avg{\gamma})$ was
observed to be linearly correlated with the percentage energy
loss of the electron beam (as compared against a reference beam,
see \cref{fig:Cole}).
Three shots were selected as exemplary cases of poor, moderate
and strong overlap, according to these two values, with
corresponding strength of radiation-reaction effects.
This shots are labelled (b), (c) and (d) respectively in
the right-hand panel of \cref{fig:Cole}.
The analysis then focussed on comparison of the measured electron energy
spectra against those predicted by simulations under various
models of radiation reaction.
These comparisons showed that the Landau-Lifshitz equation,
i.e. classical radiation reaction, overestimated the energy loss,
with a quality of fit of $R^2 = 0.87$. Simulations with the
`modified classical' model improved the agreement to $R^2 = 0.96$;
this was found to give better agreement than the fully stochastic model,
in which case $R^2 = 0.92$. This discrepancy was attributed to
possible failure of the LCFA as the collision $a_0 \simeq 10$.
Nevertheless, by considering the detailed shape of electron
energy spectra, it was possible to find clear evidence of
radiation reaction, as well as signatures of quantum corrections.

\section{Summary and outlook}

Let us now consider the relation between the results of these
two experiments discussed in \cref{sec:RecentResults}.
Both present clear evidence that radiation
reaction, in some form, has taken place. The reduction in the
electron energies, the total $\gamma$-ray signal, and, in
\citet{cole.prx.2018}, the spectral shape of the latter,
are all broadly consistent with each other.
The differences arise in the comparison of different models of
radiation reaction, bearing in mind that, in the regime
where $\chi \simeq 0.1$, $a_0 \simeq 10$, quantum corrections
are expected to be non-negligible, but not large, and
the intensity is not so large that the LCFA is beyond question.
The use of simulations that rely on this approximation is,
however, necessary because the number of photons emitted, per
electron in the beam, is much larger than unity, and therefore
an exact calculation from QED is intractable at present
(see \cref{sec:Benchmarking}).
In \citet{cole.prx.2018}, the shot-to-shot fluctuations in
the electron beam energy and alignment, and the fact
that the electron spectrum is analyzed by means of a single value
rather than its complete shape,
mean that all three models (classical, modified classical,
and quantum) are not distinguishable from each other at level
of two standard deviations. At the one-standard-deviation level,
the two models that include quantum corrections provide
better agreement.

\Citet{poder.prx.2018}, with significantly
more stable electron beams, are able to confirm that the
classical model is not consistent with the data either.
However, the fact that neither the modified classical or quantum
models provide a very good fit to the data leaves open
the question of whether it is the failure of
the LCFA or, as they state, ``incomplete knowledge
of the local properties of the laser field.''
Accurate determination of the initial conditions, in both the
electron beam and the laser pulse, will be of unquestioned
importance for upcoming experiments that aim to discern
the properties of radiation reaction in strong fields.
It will be vital to characterize the uncertainties in both
the experimental conditions and the theoretical models in our
simulations, which are inevitably based on certain approximations.

Nevertheless, these results demonstrate the capability
of currently available high-intensity lasers to probe
new physical regimes, where radiation reaction and quantum
processes become the important, if not dominant, dynamical
effects. These experiments provide vital data in the
unexplored region of parameter space $\chi \gtrsim 0.1$,
$a_0 \gg 1$ (see \cref{fig:a0-chi}), allowing us to examine
critically our theoretical and simulation approaches
to the modelling of particle dynamics in strong electromagnetic
fields. The current mismatch between simulations and
experimental data has prompted, and will continue to
prompt, new ideas in how to resolve the discrepancy:
from the development of analysis techniques that are
robust against shot-to-shot fluctuations~\citep{baird.njp.2019,arran.ppcf.2019},
to improved simulation methodologies~\citep{dipiazza.pra.2019,ilderton.pra.2019}.
These are accompanied by renewed examination of the
approximations underlying our simulations
(see \cref{sec:Benchmarking}).
The development of theoretical approaches that can go
beyond the plane-wave configuration, the background field
approximation, or low multiplicity, in strong-field QED
is vital if this theory is to be applied directly in
experimentally relevant scenarios.
There is also undoubtedly a need to gather more experimental
data and explore a wider parameter space, increasing the
electron beam energy and laser intensity, i.e. $\chi$ and $a_0$.
Not only will this make radiation reaction and quantum
corrections more distinct, it will also allow us to measure
nonlinear electron-positron pair creation by the $\gamma$ rays
emitted by the colliding electron beam~\citep{sokolov.prl.2010,%
bulanov.pra.2013,lobet.prab.2017,blackburn.pra.2017},
a strong-field QED process without classical analogue.
Such findings will underpin the study of particle and
plasma dynamics in strong electromagnetic fields for many
years to come.

\begin{acknowledgements}
I am very grateful to Arkady Gonoskov, Mattias Marklund
and Stuart Mangles for a critical reading of the manuscript.
This work was supported by the Knut and Alice Wallenberg Foundation.
\end{acknowledgements}

\bibliography{references.bib}

\end{document}